\documentclass[10pt, letter, twocolumn]{IEEEtran}

\usepackage{graphicx}
\usepackage{epsfig}
\usepackage{algorithm}
\usepackage{algorithmic}
\usepackage{ifmtarg}
\usepackage{caption}
\usepackage{footnote}
\usepackage{cite}
\usepackage{amssymb}
\usepackage{amsthm}
\usepackage[fleqn]{amsmath}
\usepackage{amsmath}
\usepackage{epstopdf}
\usepackage{amsfonts}
\usepackage{enumitem}
\usepackage{subfigure}
\usepackage{multirow}
\usepackage{rotating}
\usepackage{hyperref}
\usepackage{color}

\usepackage[font=normalsize]{caption}
\begin{document}
\title{Spatial and Temporal Management of Cellular HetNets with Multiple Solar Powered Drones}

\author{\IEEEauthorblockN{Ahmad Alsharoa, \textit{Member, IEEE}, \large Hakim Ghazzai, \textit{Member, IEEE}, Abdullah Kadri, \textit{Senior Member, IEEE}, and Ahmed E. Kamal, \textit{Fellow, IEEE}}\\
\thanks { \vspace{-0.5cm}\hrule
\vspace{0.1cm} \indent A part of this work has been accepted for publication in IEEE Wireless Communications and Networking Conference (WCNC 2017)~\cite{alsharoaWCNC}.
\newline \indent Ahmad Alsharoa is with the Engineering and Computer Computer Science Department, Virginia State University, Petersburge, Virginia 23806, USA, E-mail: aalsharoa@vsu.edu.
\newline \indent Hakim Ghazzai is with School of Systems and Enterprises, Stevens Institute of Technology, Hoboken, NJ, USA, E-mail: hghazzai@stevens.edu.
\newline \indent Abdullah Kadri is with Qatar Mobility Innovations Center (QMIC), Qatar University, Doha, Qatar. Email: abdullahk@qmic.com.
\newline \indent Ahmed E. Kamal is with the Electrical and Computer Engineering Department, Iowa State University (ISU), Ames, Iowa 50010, USA, E-mail: kamal@iastate.edu.
\newline This work of Ahmad Alsharoa was done when he was with the Electrical and Computer Engineering Department, ISU.}\vspace{-.1cm}}

\maketitle
\thispagestyle{empty}
\pagestyle{empty}

\begin{abstract}
\boldmath{
This paper proposes an energy management framework for cellular heterogeneous networks (HetNets) supported by dynamic solar powered drones. A HetNet composed of a macrocell base station (BS), micro cell BSs, and drone small cell BSs are deployed to serve the networks' subscribers. The drones can land at pre-planned locations defined by the mobile operator and at the macrocell BS site where they can charge their batteries. The objective of the framework is to jointly determine the optimal trips of the drones and the MBSs that can be safely turned off in order to minimize the total energy consumption of the network. This is done while considering the cells' capacities and the minimum receiving power guaranteeing successful communications. To do so, an integer linear programming problem is formulated and optimally solved for three cases based on the knowledge level about future renewable energy statistics of the drones. A low complex relaxed solution is also developed. Its performances are shown to be close to those of the optimal solutions. However, the gap increases as the network becomes more congested. Numerical results investigate the performance of the proposed drone-based approach and show notable improvements in terms of energy saving and network capacity.}
\end{abstract}


\section{Introduction}\label{Introduction}
\IEEEPARstart{W}ith the rapid growth of drones market, drone-based communication has been considered as a promising and effective solution to many of today's challenges in the area of wireless communications due to the low cost and fast deployment of drones, autonomous motion without human intervention, and robustness against such environmental changes such as floods and earthquakes. Recently, drones, which are also known as unmanned aerial vehicles (UAVs), are porposed to be used as small cell base stations (BSs) to support ground cellular networks has received considerable attention~\cite{surv1,surv2}. A drone BS (DBS) can act as an aerial BS characterized by a quick and dynamic deployment which is extremely helpful for various scenarios~\cite{DBSint1}. For instance, in public safety communication, where ground infrastructure is damaged by natural disasters, DBSs represent an alternative solution for mobile operators to maintain coverage and connectivity. In fact, DBSs are more robust against such environmental changes thanks to their mobility. DBSs are also useful for temporary/unexpected high traffic demand situations where already deployed infrastructure becomes overloaded and requires additional communication equipment to maintain the high quality-of-service (QoS) level. For example, in big events such as Football games, Olympic games, or Concerts, it is infeasible from economical perspective to invest in the ground infrastructure for a relatively short time period.
In this context, many companies have developed prototypes for LTE DBSs such as Nokia, AT\&T, Qulacomm, Intel~\cite{Att1,Att2}.
For instance, Nokia has showcased its newly developed LTE DBS at the UAE Drones for Good (D4G) Award event in Dubai, UAE in 2017~\cite{Nokia_ref}. This new technology provides centralized monitoring and control of DBSs via an operator’s existing LTE network or dedicated LTE network.

The placement of DBSs is considered as one of the main challenges in drone-based communications~\cite{pathdeployment1,pathdeployment3} particularly in the case of multiple DBSs. Optimizing the DBS locations can significantly enhance the network performance either by reducing the load of other ground BSs or by covering areas with limited radio access. Another challenging issue in drone-based communications is the power management of these battery limited DBSs since traditional wired charging methods are not feasible. Therefore, energy harvesting (EH) techniques can be considered as one of the most effective and robust solutions to protract the lifetime and sustainability of drones' batteries~\cite{int1}. Recently, many promising practical applications that use EH nodes have emerged such as ultra-dense small cell deployments, point-to-point sensor networks, cognitive radio networks, and far-field microwave power transfer~\cite{RFA2}. In drone-based communications, EH can also be an attractive technology to power DBSs by offering additional energy to charge their batteries\cite{UlukusEH,GCom,RE}.

\subsection{Related Work}
Few works in the literature investigated the deployment of DBSs and its challenges.
The challenges of multi-tier drone-cells in 5G radio access networks have been discussed in~\cite{H1}. A novel multi-tier drone-cells management framework has been proposed for efficient operation. The authors showed that the management framework mechanisms reduce the cost of utilizing drone-cells in multi-tenancy cellular networks.
In~\cite{relaydrone}, a placement technique that uses the drones as relays for cell overloading and outage compensation is proposed. Although an analytical model is provided for evaluating system performance in the downlink direction, the paper did not discuss the DBSs' coverage performance and did not suggest any deployment method. The authors in~\cite{relaydrone2} discussed the optimal deployment position for drones that maximizes the average data rate while keeping the symbol error rate under a certain level. However, their work is limited to only one relaying drone. In \cite{11}, the authors proposed a computational method to find the optimal and fast drone deployment in order to enhance the coverage performance in the case of public safety communications. In~\cite{H2}, Kalantari \textit{et al.} studied the backhaul resource allocation, user association, and 3D placement of drones. The authors proposed a novel delay-sensitive approach by associating the delay-sensitive users to the macro BS and the other users to either the macro BS or DBSs.
A heuristic iterative algorithm is proposed to optimize the 3D locations of the drones and the backhauling bandwidth allocation.

On the other hand, some works discussed the connectivity and safe path planing management for drone-based communication scenarios. For instance, improving the connectivity of ad-hoc networks using drones has been discussed in~\cite{14,connectivity2}. The authors in~\cite{14} developed a simple heuristic suboptimal algorithm to optimize the drones movement by tracking changes in the network. Safe path planning algorithms with multiple drones are proposed in~\cite{safepath1,safepath2} with the objective to ensure that the drones can return to the charging station before their energy is depleted.

Channel modeling in drone-based communications also remains an important research direction that has extensively been discussed~\cite{Drone_saad,LOS1,LOS2}.
In general, the ground receiver receives three groups of signals: (a) Line-of-Sight (LoS) signals, (b) strong Non Line-of-Sight (NLoS) signals such as strong reflected signals, and (c) multiple reflected components which cause multipath fading. Therefore, the aforementioned signals can be considered separately with different probabilities of occurrence. Typically, as discussed in~\cite{LOS2}, it is assumed that the received signal is categorized only in one of the mentioned groups. Each group has a specific probability of occurrence which is a function of environment, density, height of obstacles, and elevation angles.
Indeed, one of the advantages of using flying DBSs is their ability to establish the LoS link with ground users which helps in enhancing the signal quality. In~\cite{Drone_saad}, the authors analyzed the optimal altitude of one DBS for a certain coverage area that minimizes the DBS's transmit power. The work in \cite{Drone_saad} proposed to consider LoS and NLoS components along with their occurrence probabilities. Moreover, they investigated the coverage of two DBSs positioned at a fixed altitude and interfering with each other over a certain coverage area.
The probability of air-to-ground LoS link is determined in~\cite{LOS1} for a dense urban area. It depends on the altitude, elevation angle, and the distance between the drone and the user or ground node. On the other hand, the air-ground path loss (PL) model for urban environment has been discussed in \cite{LOS2}. In \cite{PL1}, the authors provided both closed-form expressions for predicted probability of LoS and PL model for air-to-ground environment using low altitude platform.
In~\cite{saad2}, the authors studied the coexistence between the drones and underlaid device-to-device (D2D) communication in the downlink scenario. More specifically, they derived the average downlink coverage probabilities for the users and analyzed the impact of the drones' altitudes and density on the overall performance for static and moving drones.

\subsection{Contributions}
In this paper, a drone-based communication problem is addressed from a new perspective by investigating the placement of multiple EH DBSs in order to support typical heterogeneous networks (HetNets) composed of a single macrocell BS and multiple ground micro cell BSs (MBSs). The proposed method can be generalized to the context of large-scale HetNets.
Several studies considering the installation of solar panels or solar films on top of the drones or aircrafts, depending on the size, have been proposed in the literature~\cite{7470932,7759260,H3,H4,quadcopter,PowerFilm}. For instance, in~\cite{PowerFilm}, it is shown that using power film can provide output around 0.72 Watt with the following power film specifications (weight: 10 oz, width: 3.5 in, length: 10.6 in, height: 0.01 in).

In this study, we assume that each drone can charge its battery either using traditional electric energy when it is placed in a charging station located at the macrocell BS site or using renewable energy (RE) harvested through solar film/panels placed on top of the drones.
The objective of the framework is to exploit the mobility and quick deployment of these solar-powered drones to support the ground cells whenever it is needed and whenever the drones' batteries permit it. Inactive drones, which are originally placed at the charging station, will be asked to fly to particular locations to serve users and support the overloaded HetNet or replace lightly loaded MBSs during a short period of time. In the latter case, the MBSs can be safely turned off to reduce fossil fuel consumption. For realistic deployment, we consider a finite  pre-planned possible locations known by the mobile operator for drone placement. At these locations, the drones can land and serve the users under their coverage. This study aims to optimize the spatial and temporal management of these multiple drones under different traffic and situations. Moreover, since the RE is random in nature, we develop a stochastic programming solution to deal with this source of uncertainty.

The main contributions of this work can be summarized as follows:
\begin{itemize}
  \item An optimization problem is formulated aiming to minimize the total fossil fuel consumption of the drones-assisted HetNet. The objective is to support the network's capacity by employing multiple drones as flying BSs to be placed at specific potential pre-planned locations.
  \item The green operation of the HetNet is investigated while taking into account several factors including a QoS metric, the cells' capacity, drones' battery limit, photovoltaic generation at the drone levels, and the power consumption related to drones' mobility.
  \item This green framework involves the application of the ON/OFF switching strategy to the MBSs whenever it is possible. A joint optimization solution is proposed for drones' placement and MBSs deactivation during a long period of time.
  \item Three cases depending on the knowledge level about future RE generation are investigated:
  \begin{enumerate}
    \item The zero knowledge case: in this case, future RE generation statistics are unknown for the mobile operator. A binary linear programming problem is formulated to determine the HetNet and drone statuses based on past and present realizations.
    \item The perfect knowledge case: this case assumes that the future statistics of the network are perfectly known and estimated. A non-linear programming problem is formulated to determine the future deployment strategies for the drones. To reduce its complexity, a linearization approach is employed to transform the problem into a binary linear programming optimization problem.
    \item The partial knowledge case: in this case, only partial statistics of the future RE generation are known, i.e., probability density function. To deal with the uncertainty effect, a stochastic programming problem is formulated and solved using the two-stage recourse method. In this case, the uncertainty effect is also considered.
  \end{enumerate}
  \item A relaxed solution is proposed based on subgradient algorithm to find near optimal solution with low complexity.
\end{itemize}
Numerical results are provided to demonstrate the role of drones in supporting the HetNet. They can either replace redundant MBSs if their energy consumption is lower than that of the MBS or support the existing infrastructure to increase the network capacity. In addition, numerical results offer a comprehensive comparison of the different cases and analyze the effect of RE generation uncertainty on the energy consumption of the network.

\subsection{Paper Organization}
The remainder of the paper is organized as follows. Section~\ref{SystemModel} presents the drone-based communication system model. The problem formulations and the corresponding solutions are given in Section~\ref{ProblemFormulation}. Section~\ref{Simulations} introduces and discusses some numerical results. Finally, the paper is concluded in Section~\ref{Conclusions}.

\section{System Model}\label{SystemModel}
In this study, we investigate a time-slotted system of a finite period of time divided into $B$ slots numbered $b=1,\cdots,B$, of equal duration $T_b$. Investigating the system performance for instantaneous channel realizations and network statistics is not valid for this framework since we are considering the drones' flying time (seconds) which is very large as compared to the channel coherence time (milliseconds).

\subsection{Network Model}
\begin{figure}[t!]
  \centerline{\includegraphics[width=3.5in]{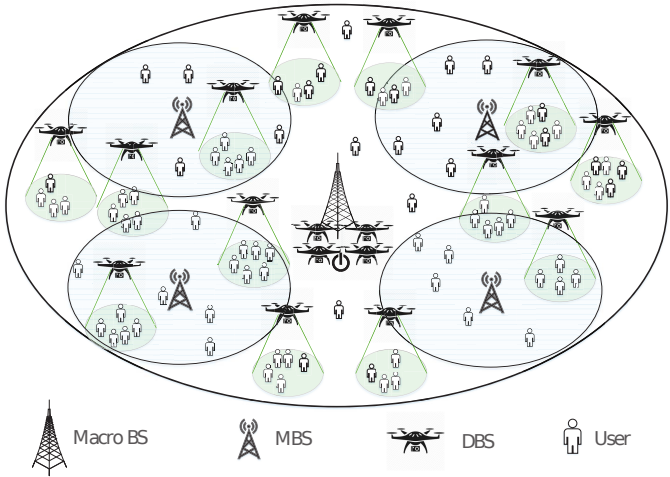}}
   \caption{\small Example of a HetNet assisted by DBSs. \normalsize}\label{fig0}
\end{figure}

We consider a typical HetNet consisting of one macrocell BS and $M$ MBSs. The HetNet is assisted by $D$ dynamic drones that act as DBSs (i.e., DBS is carried by one drone) as depicted in Fig.~\ref{fig0}. For the readers' convenience, the symbols and notations used in this paper are summarized in Table~\ref{Tab3}.

In this work, we aim to optimize the deployment of DBSs in the geographical area covered by the macrocell BS according to the network's need and QoS requirements. We assume that a dynamic drone can be in three different states: 1) the drone is in an idle mode and placed at the charging station assumed to be located in the center of the cell (i.e., in the macrocell BS site.), 2) the drone is placed at a pre-defined location in the cell and acting as a DBS to serve users, or 3) the drone is in motion and flying from a location to another. Placing the charging station at the center of the cell minimizes, in general, the flying time of drones and hence the corresponding energy consumption.

In Table~\ref{Tab3}, we summarize the notations used in this paper.
\begin{table}[h!]
\begin{center}
\caption{List of Notations}
\label{Tab3}
\begin{tabular}{|c|c|}%
\hline
   \textbf{Notation} & \textbf{Description}\\
\hline
$M$ & 	Number of MBS  \\
\hline
  $D$  & Number of drones  \\
\hline
$Z$  & Number of drones' possible locations \\
\hline
$B$  & Number of time slots \\
\hline
$x_i,y_i,h_i$ &  drone 3D geographical coordinates \\
\hline
 $\mathbf{\epsilon^b}$ & A binary matrix indicates the drones locations \\& for a given $b$\\
\hline
 $\mathbf{\pi^b}$ & A binary vector indicates the MBS status \\
 \hline
$U^b$  &  The total number of users during time slot $b$ \\
\hline
 $\bar{U}_0,\bar{U}_m,\bar{U}_d$  & The maximum number of users \\& served by macrocell BS, MBS $m$, and DBS $d_l$ \\
\hline
$\mathcal{A}_\mathcal{X}$  & The coverage area of an active BS $\mathcal{X}$ \\
\hline
 $PL_{\mathcal{X}}^\text{NLoS},PL_{\mathcal{X}}^\text{LoS}$  & The path loss of NLoS and LoS, respectively \\
\hline
$\lambda_0$  &  The carrier wavelength \\
\hline
$\xi_{\text{NLoS}},\xi_{\text{LoS}}$  & The average propagation loss of NLoS \\& and LoS, respectively \\
\hline
$\alpha_\mathcal{X}$  & The scaling power of BS $\mathcal{X}$ \\
\hline
$\beta_\mathcal{X}$  & The offset site power of BS $\mathcal{X}$ \\
\hline
$\gamma_\mathcal{X}$  &  Minimum power required to readily activate BS $\mathcal{X}$ \\
\hline
$v_d,v_\text{max}$  & Drone speed, maximum drone speed, respectively \\
\hline
$P_\text{hov}$  &  The drone's hovering power \\
\hline
$P_\text{har}$  &  The drone's hardware power \\
\hline
$P_\text{full}, P_\text{s}$  &   The drone hardware power at full speed\\& and in idle mode, respectively\\
\hline
$P_f$  & The consumed drone flying power\\
\hline
$P_{ch}$  & The charging power at charging station. \\
\hline
$m_\text{tot}$  & The drone's mass\\
\hline
$g, \rho$  & The Earth gravity and air density, respectively\\
\hline
$r_p$  & The drone’s propellers radius\\
\hline
$n_p$  & The number of the drone’s propellers\\
\hline
$T_b,T_f$  & Slot and flying times, respectively \\
\hline
$E^b_0,E^b_M,E^b_D$  & The energy consumption of the macrocell BS,\\& MBS, and DBS, respectively, for a given $b$ \\
\hline
$S_{d_l}^b, H_{d_l}^b$  & The stored and harvest energies at DBS $d_l$,\\& respectively, at the end of time slot $b$\\
  \hline
  \end{tabular}
\end{center}
\end{table}

We assume that there are $Z+1$ possible locations available for drones' deployment. These locations, $i=0,\cdots,Z$, can be pre-determined by the mobile operator during the planning phase depending on several factors such as, historical network statistics, location constrains, etc. Each location $i$ is identified by its three dimensional (3D) geographical coordinates $(x_i,y_i,h_i)$. The location $i=0$ (i.e., $x_0=y_0=h_0=0$) corresponds to the charging station. Hence, the drone energy consumption depends essentially on its current location (i.e., time slot $b$) and the previous one (i.e., time slot $b-1$). Note that once the drone landed in a specific location, it will be in a static state instead of a moving state. This can help in reducing the energy consumption of the drones and improve the battery lifetime.

We denote by $\boldsymbol{\epsilon^b}$ a binary matrix of size $D \times (Z+1)$. Its entries $\epsilon_{d_l}^{b}(i)$ indicate the location of the drone $d_l$, where $l=1,\cdots, D$, and is given by
\begin{equation}
   \epsilon^b_{d_l}(i)=\left\{
   \begin{array}{ll}
   1, & \hbox{if the drone $d_l$ is placed at location $i$} \\
	 & \hbox{during time slot $b$},\\
   0, & \hbox{otherwise.}
                   \end{array}
                 \right.
\end{equation}

On the other hand, a dynamic ON/OFF switching mechanism is considered to turn off redundant MBSs whenever it is possible~\cite{RR7,7776901}. More specifically, MBS $m_k, k=1,\dots, M$ can be turned off during low traffic periods and the small number of active users are offloaded to nearby DBSs or macrocell BS.
Note that the duration time of turning off BSs overhead is very small compared to the length of the time slots $T_b$. Therefore, this overhead time is neglected.
As a result, the energy consumption of lightly loaded MBSs can be eliminated. A binary vector $\boldsymbol{\pi}^b$ is introduced to indicate the status of each ground MBS $m_k$ and is given by:
\begin{equation}
  \pi^b_{m_k}=\left\{
   \begin{array}{ll}
   1, & \hbox{if MBS $m_k$ is operating during time slot $b$.} \\
   0, & \hbox{otherwise.}
                   \end{array}
                 \right.
\end{equation}

It should be noted that we always keep the macrocell BS active to ensure coverage and minimum connectivity in this typical HetNet (i.e., one macrocell BS surrounded by multiple of MBSs). In the case of multiple macrocell BSs covering a bigger geographical area, macrocell BSs could be turned off and cell breathing mechanisms can be employed to ensure connectivity~\cite{GCom}.

We denote by $U^b$ the average total number of users located in the macrocell BS during time slot $b$ and by $\bar{U}_0$, $\bar{U}_m$, and $\bar{U}_d$ the maximum number of users that can be served by macrocell BS, MBS $m_k$, and DBS $d_l$, respectively, such that $\bar{U}_d \leq \bar{U}_m \ll \bar{U}_0$. These numbers reflect the BSs' capacities due to available number of frequency carriers and/or hardware and transmit power limitations. We assume that the co-channel interference is ignored and the transmissions are performed in orthogonal basis. This is not a hard assumption since the time slots are relatively long compared to the channel coherence time and hence, we focus on the average statistics of the network. This implies that average interference might be assumed or mitigated through frequency re-use techniques. Hence, the interference effect becomes a non-challenging issue which does not impact the problem formulation. Also, we assume that a user is served by at most one BS (either a macrocell BS, MBS, or DBS). We consider that the user distribution during time slot $b$ over the macrocell BS area $\mathcal{A}$ follow a certain probability density function (pdf) denoted by $f(x,y,b)$, where $(x,y)$ represents the geographical coordinates of a user. We denote by $\mathcal{A}_\mathcal{X}$ $(\mathcal{A}_\mathcal{X} \subseteq \mathcal{A})$ the coverage area of an active BS $\mathcal{X}$ where $\mathcal{X} \in \{\{0\}, \{m_k: k=1,\cdots, M\}, \{(d_l,i):l=1,\cdots,D,i=0,\cdots,Z\}\}$ referring to the macrocell BS, MBS $m_k$, and DBS $d_l$ placed at location $i$, respectively. Hence, the average number of users served by an active BS $\mathcal{X}$ during time slot $b$ is denoted by $U^b_{\mathcal{X}}$ and is given by:

\small
\begin{align}
&\hspace{-0.5cm}U^b_{\mathcal{X}}=\min \left(U^b \iint_{\mathcal{A}_\mathcal{X}} f(x,y,b) \ \mathrm{d}x \mathrm{d}y, \bar{U}_\mathcal{X}\right),\\
&\,\text{    for MBSs and DBSs, i.e., $\mathcal{X} \in \{m_k, (d_l,i)\}$ and,} \notag\\
&\small \hspace{-0.5cm}U^b_0=\min \left(U^b - \sum \limits_{k=1}^M \pi^b_{m_k} U^b_{m_k}- \sum \limits_{l=1}^D \sum \limits_{i=1}^Z \epsilon^b_{d_l}(i) U^b_{d_l,i} , \bar{U}_0
\right),\label{MacroUsers}\\
& \text{ for macrocell BS, i.e, $\mathcal{X}= {0}$.}\notag
\end{align}\normalsize
where $\min(.,.)$ is the minimum function. In \eqref{MacroUsers}, the priority in serving users is given to MBSs and DBSs as macrocell BS's transmit power is usually higher than that of MBS and DBS~\cite{EARTH}.

\subsection{Path Loss Model}
In this paper, we are dealing with the average network statistics. Therefore, the channel model is essentially based on the PL model. Fast-fading effects are ignored in this study. Two PL models can be distinguished depending on the nature of the transmitter. The PL model employed in the present paper is used in many previous studies on UAV-based communications, e.g., \cite{Drone_saad,Saaad,EH_drone,alsharoaICC}.
\subsubsection{Ground-to-Ground Path Loss Model}
The average PL between a ground BS $\mathcal{X} \in \{0,m_k\}$ and a ground user is given by the average PL for the NLoS link and is expressed by~\cite{PL1}:
\begin{equation}\label{PLnlos}
PL_{\mathcal{X}}^\text{NLoS} [\text{dB}]= 20 \log_{10}\left(\frac{4 \pi \delta_\mathcal{X}}{\lambda_0}\right)+\xi_{\text{NLoS}},
\end{equation}
where $\delta_\mathcal{X}$ is the average distance between the ground BS $\mathcal{X}$ and a served user located within its cell, $\lambda_0$ is the carrier wavelength, and $\xi_{\text{NLoS}}$ is the average additional loss due to the free space propagation loss for NLoS link.

\subsubsection{Air-to-Ground Path Loss Model}
The PL of the air-to-ground link is a weighted combination of two PL links: LoS and NLoS links. This is due to the ability of the drones to serve users from high altitude as compared to ground BSs. In this case, there will be a probability to obtain a LoS link between the DBS and a user~\cite{PL1}. The average PL between the DBS $l$ positioned at a position $i$ and a served user in urban environments for LoS link is given as~\cite{PL1}:
\begin{equation}
PL^{\text{LoS}}_{d_l,i} [\text{dB}]= 20 \log_{10}\left(\frac{4 \pi \delta_{d_l,i}}{\lambda_0}\right)+\xi_{\text{LoS}},
\end{equation}
where $\delta_{d_l,i}$ is the average distance between the DBS $l$ located at position $i$ and the served user located in its cell and $\xi_{\text{LoS}}$ is the average additional loss to the free space propagation loss for LoS link.

The LoS probability is given by~\cite{PL2,PL3,Drone_saad}:
\begin{equation}
p^{\text{LoS}}_{d_l,i}=\frac{1}{1+\nu_1 \exp(-\nu_2[\theta_{d_l,i}-\nu_1])},
\end{equation}
where $\theta_{d_l,i}$ is the elevation angle between DBS $l$ positioned at location $i$ and the served user in degree. $\nu_1$ and $\nu_2$ are constant values that depend on the environment. The NLoS probability is, then, equal to $1-p^{\text{LoS}}_{d_l,i}$. Therefore, the average PL for air-to-ground link is given by:
\begin{equation}\label{avgPLd}
\overline{PL}_{d_l,i}=p^{\text{LoS}}_{d_l,i}PL^{\text{LoS}}_{d_l,i}+(1-p^{\text{LoS}}_{d_l,i}) PL^\text{NLoS}_{d_l,i}.
\end{equation}

\subsection{Base Stations Power Model}
In the active state and to serve its connected users during a time slot $b$, the BS $\mathcal{X}$ consumes power denoted by $P_{\mathcal{X}}^b$. However, in the idle mode, it consumes a constant power equal to $ P_{\mathcal{X}}^{\text{idle}}= \gamma_{\mathcal{X}}$ so it can be quickly reactivated.
For simplicity, the total power consumption of an active BS $\mathcal{X}$ during a time slot $b$ can be approximated by a linear model as follows~\cite{EARTH}:
\begin{equation}\label{BSpowermodel}
  P_{\mathcal{X}}^b= \alpha_{\mathcal{X}} \tilde{P}_{\mathcal{X}}^b+\beta_{\mathcal{X}},
\end{equation}
where $\alpha_{\mathcal{X}}$ is a parameter that scales with the radiated power, denoted by $\tilde{P}_{\mathcal{X}}^b$, and $\beta_{\mathcal{X}}$ models constant power.
The radiated power of a BS $\mathcal{X}$ is expressed~as:
\begin{equation}\label{radiated_power}
\tilde{P}_{\mathcal{X}}^b=U^b_{\mathcal{X}} P_{\min} {\overline{PL}}_{\mathcal{X}},
\end{equation}
where ${\overline{PL}}_{\mathcal{X}}$ is the corresponding average PL at the BS cell $\mathcal{X}$. Note that  ${\overline{PL}}_{\mathcal{X}}=PL^{\text{NLoS}}_{\mathcal{X}}$ given in~\eqref{PLnlos} in the case of a macrocell BS or MBS and ${\overline{PL}}_{\mathcal{X}}={\overline{PL}}_{d_l,i}$ given in~\eqref{avgPLd} in the case of a DBS.

\subsection{Drone Power Model}
Besides the power consumed by the BSs carried by the drones (i.e., DBSs), the drone consumes additional hovering and hardware powers. Without loss of generality, we assume that all drones move with a fixed speed denoted by $v_d$. The hover and hardware power levels, denoted by $P_\text{hov}$ and $P_\text{har}$,  can be expressed, respectively, as~\cite{Dpower_model}:
\begin{equation}\label{Ps}
P_\text{hov}= \sqrt{\frac{(m_\text{tot} g)^3}{2 \pi r_p^2 n_p \rho}}, \text{ and } P_\text{har}= \frac{P_\text{full}-P_s}{v_\text{max}}v_d+P_s,
\end{equation}
where $m_\text{tot}$, $g$, and $\rho$ are the drone mass in ($\text{Kg}$), earth gravity in (m$/\text{s}^2$), and air density in $(\text{Kg}/\text{m}^3)$, respectively. $r_p$ and $n_p$ are the radius and the number of the drone's propellers, respectively. $v_\text{max}$ is the maximum speed of the drone. $P_\text{full}$ and $P_s$ are the hardware power levels when the drone is moving at full speed and when the drone is in idle mode, respectively. Note that, in~\eqref{Ps}, we assume that when serving users at a location $i$, the drone will be in a static position, hence, it consumes only $P_s$ for hardware power. However, when it is flying to a destination (i.e., one of the $Z+1$ locations), it will consume $P_\text{har}$. Finally, the flying power of DBS $l$ can be calculated as:
\begin{equation}
P_f= P_\text{hov}+P_\text{har}.
\end{equation}
\subsection{Renewable Energy Model}
In this paper, we assume that DBS $l$ can harvest from RE sources selected to be the photovoltaic energy. We model the RE stochastic energy arrival rate as a random variable $\Phi$ Watt defined by a pdf $f_\Phi(\varphi_{d_l}^b)$~\cite{stochastic_book}. An event $\eta\varphi_{d_l}^b$ in a time slot $b$ can be interpreted as the average received amount of power with respect to the received luminous intensity in a particular direction per unit solid angle. The parameter $\eta$ denotes the EH efficiency coefficient.

\begin{table*}
\vspace{-2mm}
\begin{tabular}{l}
\begin{minipage}[b]{0.9775\textwidth}
\centering
\caption{Consumed and harvested energies of DBS $l$ during a time slot $b$ for all possible cases}\vspace{-0.2cm}
\label{Tab1}
\addtolength{\tabcolsep}{-4pt} \small\begin{tabular}{|c|c|c|c|c|c|}
\hline
Case & Previous location& Current location&Consumed energy & Harvested energy & Charging energy \\ \hline
(1)& $\epsilon^{b-1}_{d_l}(j)=1$,  $j\neq i$ & $\epsilon^{b}_{d_l}(i)=1$, $i\neq 0$   & $(P_f+ \gamma_d)T_f(j,i)+(P^b_{d_l,i}+P_s)T_r(j,i)$ &
$\eta\varphi_{d_l}^b (T_f(j,i)+T_r(j,i))$   & $0$     \\ \hline
(2)& $\epsilon^{b-1}_{d_l}(j)=1$,  $j=i$     &  $\epsilon^{b}_{d_l}(i)=1$, $i\neq 0$   & $(P_{d_l,i}^b+P_s)T_b$                    &
$\eta\varphi_{d_l}^bT_b$                    &  $0$     \\ \hline
(3)& $\epsilon^{b-1}_{d_l}(j)=1$,  $j\neq 0$ & $\epsilon^{b}_{d_l}(i)=1$, $i=0$       & $(P_f+\gamma_d) T_f(j,i)+\gamma_d T_r(j,i)$  &
$\eta\varphi_{d_l}^b (T_f(j,i)+T_r(j,i))$   & $P_{ch} T_r(j,i)$    \\ \hline
(4)& $\epsilon^{b-1}_{d_l}(j)=1$,  $j=0$     & $\epsilon^{b}_{d_l}(i)=1$, $i=0$       & $\gamma_d T_b$                    &
$\eta\varphi_{d_l}^bT_b$                    & $P_{ch} T_b$        \\ \hline
\end{tabular}
\vspace{-2pt}
\end{minipage}
\end{tabular}
\vspace{-15pt}
\end{table*}
In general, the RE generation matrix $\boldsymbol{\Phi}$, of size $D \times B$, with elements $\varphi_{d_l}^b, \forall l=1,\cdots,D, \forall b=1,\cdots,B$ can be modeled as follows:
\begin{equation}\label{Phisto}
\boldsymbol{\Phi}=\bar{\Phi}+\tilde{\Phi},
\end{equation}
where $\bar{\Phi}$ is the deterministic portion of the RE generation that can be estimated from historical data and $\tilde{\Phi}$ is a matrix of random variables representing the stochastic portion of the RE generation and models its uncertainty.

To summarize, we present in Table~\ref{Tab1} the consumed and harvested energies of the drone for all possible cases: 1) when the drone is moving from a location $j\neq 0$ to a new location $i\neq 0$, 2) when the drone remains at the same location $i\neq 0$, 3) when the drone decides to go to the charging station (i.e., $i=0$) while it was positioned at location $j\neq 0$ during time slot $b-1$, and 4) when the drone decides to stay in the charging station $i=0$. In Table~\ref{Tab1}, $T_f(j,i)$ corresponds to the drone trip duration from a location $j$ to a location $i$ and it is computed as follows:
\begin{equation}
T_f(j,i)=\frac{d_{j,i}}{v_d},
\end{equation}
where $d_{j,i}$ is the euclidean distance separating the two locations $i$ and $j$. On the other hand, $T_r(j,i)=T_b-T_f(j,i)$ corresponds to the time spent by a drone at a location $i$ to serve users (i.e., $i\neq 0$) or to charge its battery (i.e., $i=0$) such that $T_f(j,i)\ll T_r(j,i)$. $P_{ch}$ denotes the charging power per drone of the charging station.

\section{Problem Formulation}\label{ProblemFormulation}
In this section, we formulate three optimization problems, based on the knowledge level of the RE generation, aiming to minimize the network's energy consumption during the $B$ time slots. Choosing this metric reduces at maximum the use of drones and hence, sends them only when needed. Please note that, the formulated optmization problems are not only focuse on energy saving, but also consider supporting the users and avoiding the overloaded risk as much as they can. Please note that, the formulated optimization problems are not only focused on energy saving, but also consider supporting the users and avoiding the overloaded risk as much as they can. Based on the system parameters, the optimizer will determine whether using drones would be more energy efficient than using ground infrastructure or not. For lightly loaded networks, the optimizer may decide to use drones instead of micro cell BS. This depends on many factors such as the locations of the users to be served. For instance, if they are very far, the drone will consume higher hover and transition powers. Hence, the optimizer may decide not to use the drones. In the case of highly loaded networks, the optimizer is forced to use the network full capacity including the drones independently of the users' locations and their energy consumption in order to meet its demand. The optimization problem that we formulated will decide how many drones to be used such the needs that the network are satisfied.

The first optimization problem corresponds to the zero knowledge case where the mobile operator manages its BSs time slot by time slot without any prior information about the future RE generation.
The second one corresponds to the perfect knowledge case with full information about the future RE generation where all the decision variables are simultaneously optimized for the $B$ time slots. The perfect knowledge case is a not realistic case. In this study, it is used as a benchmark scenario for comparison with other cases or as an approximation of the case where RE energy uncertainty is close to negligible. The third one assumes the availability of statistical information about the future RE generation. Hence, the network's management based on RE uncertainty will be investigated under this partial knowledge case.

In general, the total energy consumption of the network during time slot $b$ can be expressed as
\begin{equation}
E^b_{\text{tot}}= E^b_0+E^b_M+ E^b_D,
\end{equation}
where, using~\eqref{MacroUsers} and~\eqref{BSpowermodel}, $E^b_0=\left(\alpha_0\tilde{P}^b_{0}\left(\boldsymbol{\pi}^b,\boldsymbol{\epsilon}^b\right)+\beta_0\right)T_b$ and represents the energy consumption of the macrocell BS during time slot $b$. $E^b_M$ is the total energy consumption of $M$ MBSs during time slot $b$ which is expressed as:
\begin{equation}
E^b_M=\sum \limits_{k=1}^M \left[\pi^b_{m_k}(\alpha_{m} \tilde{P}_{m_k}^b+\beta_{m})+(1-\pi^b_{m_k})\gamma_{m} \right]T_b.
\end{equation}
Finally, $E^b_D=\sum_{l=1}^{D}E^b_{d_l}$ corresponds to the total energy consumption of all $D$ drones during time slot $b$. Using Table~\ref{Tab1} and knowing that $T_f(i,i)=0$, the total energy consumption of a drone $d_l$ during time slot $b$ is expressed as follows:

\small
\begin{equation}\label{consumption}
\begin{split}
&E_{d_l}^b=\epsilon^{b}_{d_l}(0)\sum \limits_{j=0}^Z \epsilon^{b-1}_{d_l}(j) \left[(P_f+\gamma_d) T_f(j,0)+\gamma_d T_r(j,0)\right]+\\&
\sum \limits_{i=1}^Z \sum_{j=0}^Z \epsilon^{b}_{d_l}(i) \epsilon^{b-1}_{d_l}(j) \left[(P_f+ \gamma_d)T_f(j,i)+(P_{d_l}+P_s)T_r(j,i)\right].
\end{split}
\end{equation}
\normalsize
On the other hand and again using Table~\ref{Tab1}, the total harvest-plus-charging energy of DBS $l$ during time slot $b$ due to EH and $P_{ch}$, denoted by $H_{d_l}^b$, is given as follows:

\small
\begin{equation}\label{harvesting}
\begin{split}
&H_{d_l}^b=\epsilon^{b}_{d_l}(0)\sum \limits_{j=0}^Z  \epsilon^{b-1}_{d_l}(j)\left[\eta\varphi_{d_l}^b T_f(j,0)+(\eta\varphi_{d_l}^b+P_{ch})T_r(j,0)\right]\\&
+\sum \limits_{i=1}^Z \sum_{j=0}^Z \epsilon^{b}_{d_l}(i) \epsilon^{b-1}_{d_l}(j) \eta\varphi_{d_l}^b \left[T_b\right].
\end{split}
\end{equation}
\normalsize

We assume that the DBSs are battery powered devices. Therefore, the stored energy by DBS $l$ at the end of time slot $b$, denoted by $S_{d_l}^b$, is given by:
\begin{equation}\label{storingexpression}
S_{d_l}^b=S_{d_l}^{b-1}+ H_{d_l}^b-E_{d_l}^b.
\end{equation}
We assume that, initially, each battery is charged by an amount of energy denoted by $S_{d_l}^0$.
We assume that the consumed energy due to signaling and computation is included in the fixed parameters of the energy models ($\beta_{\mathcal{X}}$ and $\gamma_{\mathcal{X}}$) of the different types of base stations (macrocell, MBSs, and DBSs). In this paper, we are focusing on energy consumption levels due to long term operation of the network rather than the signaling part. This is due to the fact, that the signaling is happening for very short periods of time (i.e, of the order of milliseconds) compared to longer periods of the network operation (i.e., of the order of minutes).

\subsection{Zero Knowledge Case}
In this case, we assume that the mobile operator is not aware of the future RE generation process, i.e., $\varphi_{d_l}^b$ is unknown during any future time slots, where $\boldsymbol{\Phi}$ in \eqref{Phisto} is known during $b$ only.

The optimization problem minimizing the total energy consumption at each time slot $b$ with EH drones is given~as:

\begin{align}
&\hspace{-0.5cm}\underset{\boldsymbol{\epsilon}^b \in\{0,1\}, \boldsymbol{\pi}^b\in\{0,1\}}{\text{minimize}} \quad  \quad
E^b_{\text{tot}}= E^b_0+E^b_M+ E^b_D      \label{of}\\
&\hspace{-0.5cm}\text{subject to:}\nonumber\\
&\hspace{-0.5cm}E_{d_l}^b \leq  S_{d_l}^{b-1}, \quad \forall l,\label{consuming}\\
&\hspace{-0.5cm}S_{d_l}^{b-1}+ H_{d_l}^b\leq \bar{S}, \quad \forall l,\label{storing}\\
&\hspace{-0.5cm}\sum \limits_{i=0}^Z \epsilon^{b}_{d_l}(i)=1, \quad \forall l,\label{epsilonDrone}\\
&\hspace{-0.5cm}\sum \limits_{l=1}^D \epsilon^{b}_{d_l}(i)\leq 1, \quad \forall i=1,\cdots,Z,\label{epsilonLocation}\\
&\hspace{-0.5cm}U^b - \sum \limits_{k=1}^M \pi^b_{m_k} U^b_{m_k}- \sum \limits_{l=1}^D \sum \limits_{i=1}^Z \epsilon^b_{d_l}(i) U^b_{d_l,i} \leq \bar{U}_0.     \label{capacity0}
\end{align}
Constraint~\eqref{consuming} indicates that the total energy consumed by a drone $d_l$ during the time slot $b$ has to be less than the energy stored at the beginning of this time slot. Constraint~\eqref{storing} forces the total energy stored in the battery of a drone $d_l$ during the time slot $b$ to be less than the battery capacity denoted by $\bar{S}$. Note that $\bar{S}$ is chosen such that the required energy to return a drone to the charging station ($i=0$) is guaranteed. This energy is simply equal to $P_fT_f(i_{\text{max}},0)$ where $i_{\text{max}}$ is the farthest location from $i=0$. Constraints~\eqref{epsilonDrone} and~\eqref{epsilonLocation} prevent the optimization problem from positioning a drone in two or more different locations during the same time slot and positioning at maximum one drone in the locations $i=1,\cdots, Z$, respectively. Note that multiple drones can be located simultaneously at the charging station $i=0$. Finally, constraint~\eqref{capacity0} ensures that the macrocell BS's capacity is not violated. This constraint encourages the activation of MBSs and the deployment of DBSs during high traffic time slots.

Notice that this optimization problem will be solved at the beginning of each time slot which is possible due to the knowledge of the status of the network during the previous time slot $\boldsymbol{\epsilon^{b-1}}$. Hence, the problem can be converted to the standard form of a binary linear programming optimization problem. Optimal solutions for such a problem can be determined using Gurobi/CVX interface~\cite{Gurobi}.

\subsection{Perfect Knowledge Case}
In this case, we assume that the mobile operator can perfectly predict the future RE generation $\varphi_{d_l}^b, \forall l=1,\cdots,D, \forall b=1,\cdots,B$, ahead of time (i.e., $\tilde{\Phi}=0$).
This case can be considered as a useful benchmark to compare with other cases.
Therefore, the objective function becomes the minimization of the total energy consumption of the network during all $B$ time slots. The decision variables are identified as $\boldsymbol{\epsilon}$ and $\boldsymbol{\pi}$ that correspond to the vertical concatenation of the matrices $\boldsymbol{\epsilon}^b$ and $\boldsymbol{\pi}^b, \forall b=1,\cdots, B$, respectively. Hence, the problem becomes a binary non-linear programming problem due to the existence of the binary products $\epsilon^{b-1}_{d_l}(j)\epsilon^{b}_{d_l}(i)$ in the energy expressions given in~\eqref{consumption} and~\eqref{harvesting}. To linearize the problem, we introduce for each link the parameter $\zeta^{b}_{d_l}(j,i)$ such that $\zeta^{b}_{d_l}(j,i)=\epsilon^{b-1}_{d_l}(j)\epsilon^{b}_{d_l}(i)$ where the following inequalities have to be respected:
\begin{align}
&\zeta^{b}_{d_l}(j,i)\leq \epsilon^{b-1}_{d_l}(j),\quad \, \zeta^{b}_{d_l}(j,i)\leq \epsilon^{b}_{d_l}(i),\notag\\
&\text{ and }\zeta^{b}_{d_l}(j,i)\geq \epsilon^{b-1}_{d_l}(j)+\epsilon^{b}_{d_l}(i)-1.~\label{Linear}
\end{align}
The first two inequalities ensure that $\zeta^{b}_{d_l}(j,i)=0$ if $\epsilon^{b-1}_{d_l}(j)$ or $\epsilon^{b}_{d_l}(i)$ is zero. The third inequality guarantees that $\zeta^{b}_{d_l}(j,i)=1$ if $\epsilon^{b-1}_{d_l}(j)=\epsilon^{b}_{d_l}(i)=1$.
It can be deduced from~\eqref{Linear} that when $\zeta^{b}_{d_l}(j,i)=1$, the drone $d_l$ will move from location $j$ to location $i$ during time slot $b$.
Hence, the expressions~\eqref{consumption} and~\eqref{harvesting} become depending on $\zeta_{b}(d_l,n)$ and the decision variables turn into $\boldsymbol{\zeta}$, $\boldsymbol{\epsilon}$, and $\boldsymbol{\pi}$ that have the following number of elements: $BD{(Z+1)}^2$, $BD(Z+1)$, and $BM$, respectively. Accordingly, the optimization problem that minimizes the network energy consumption during all $B$ time slots is given~by
\begin{align}
&\hspace{-0.5cm}\underset{\underset{\boldsymbol{\zeta} \in \{0,1\}}{\boldsymbol{\epsilon} \in \{0,1\}, \boldsymbol{\pi} \in \{0,1\},}}{\text{minimize}} \quad  \quad
E_{\text{tot}}=\sum \limits_{b=1}^B E^b_0+E^b_M+ E^b_D      \label{ofL}\\
&\hspace{-0.5cm}\text{subject to:}\nonumber\\
&\hspace{-0.5cm} \sum_{t=1}^{b} E_{d_l}^t-\sum_{t=1}^{b-1} H_{d_l}^t \leq  S_{d_l}^{0}, \quad \forall l, \forall b,\label{consumingL}\\
&\hspace{-0.5cm}  S_{d_l}^{0}+\sum_{t=1}^{b} H_{d_l}^t-\sum_{t=1}^{b-1} E_{d_l}^t \leq  \bar{S},  \forall l, \forall b, \label{storingL}\\
&\hspace{-0.5cm}\sum \limits_{i=0}^Z \epsilon^{b}_{d_l}(i)=1, \quad \forall l, \forall b,\label{epsilonDroneL}\\
&\hspace{-0.5cm}\sum \limits_{l=1}^D \epsilon^{b}_{d_l}(i)\leq 1, \quad \forall i=1,\cdots,Z, \forall b,\label{epsilonLocationL}\\
&\hspace{-0.5cm}U^b - \sum \limits_{k=1}^M \pi^b_{m_k} U^b_{m_k}- \sum \limits_{l=1}^D \sum \limits_{i=1}^Z \epsilon^b_{d_l}(i) U^b_{d_l,i} \leq \bar{U}_0, \quad \forall  b,      \label{capacity0L}\\
&\hspace{-0.5cm}\zeta^{b}_{d_l}(j,i) \leq \epsilon^b_{d_l}(i),\quad \forall l, \forall i, \forall j, \forall b,\label{zeta1}\\
&\hspace{-0.5cm}\zeta^{b}_{d_l}(j,i) \leq \epsilon^{b-1}_{d_l}(j),\quad \forall l, \forall i, \forall j, \forall b,\label{zeta2}\\
&\hspace{-0.5cm}\zeta^{b}_{d_l}(j,i) \geq \epsilon^{b-1}_{d_l}(j)+\epsilon^b_{d_l}(i)-1,\quad \forall l,\forall i, \forall j,\forall b,\label{zeta3}
\end{align}
Notice that the constraints~\eqref{consumingL}-\eqref{capacity0L} are similar to the constraints~\eqref{consuming}-\eqref{capacity0} except that they have to be satisfied for all time slots $b=1,\cdots, B$. The constraints~\eqref{consumingL}-\eqref{storingL} are obtained by replacing $S_{d_l}^b$ by its expression given in~\eqref{storingexpression}. The constraints~\eqref{zeta1}-\eqref{zeta3} correspond to the linearization process as indicated in~\eqref{Linear}. In terms of complexity, the linearized perfect knowledge optimization problem is largely more complex than the one of the zero knowledge case due to the higher number of binary decision variables and constraints. The linearized perfect knowledge problem can be also solved using Gurobi/CVX interface~\cite{Gurobi}.

\subsection{Partial Knowledge Case}
In this case, we assume that the mobile operator has only partial knowledge about future RE generation (i.e., the RE generation is uncertain).
One of the ways to deal with the RE uncertainty is to solve the optimization problem using stochastic programming. Stochastic programming is a mathematical framework for modeling optimization problems in which some or all optimization variables are presented by random variables that involve uncertainty.
The goal of such framework is to provide useful analytical or numerical information to a decision maker by finding a feasible policy that optimizes the expectation of some functions of the deterministic and the random decision variables~\cite{stochastic_book}. In this paper, we use two-stage recourse stochastic programming to represent the impacts of uncertainty in the partial knowledge case due to its simplicity. This approach includes two stages. In the first stage, the decision is made before observing the stochastic variables. Once the uncertain events have been unfolded, further decision on the operation of the system can be made through the second stage~\cite{zalloumi}. The first stage in stochastic programming is to optimize other variables, given that the output variables are known, for any given value of $\Phi$. Then, the decision needs to be updated once the actual realization of $\Phi$ has been obtained. More specifically, we choose to fix feasible values of $\boldsymbol{\pi}$ since it does not directly depend on the RE generation. This allows us to compute the best combination of other variables (i.e., $\boldsymbol{\zeta}$ and $\boldsymbol{\epsilon}$) provided that $\Phi$ is known.

The objective function given in~\eqref{ofL} can be re-written as $E_{\text{tot}}=\hat{E}+E^*$, where $\hat{E}$ does not depend on the RE directly while $E^*$ does. Therefore, the problem can be written as a two-stage recourse problem as follows~\cite{tss}:
\begin{equation}
\underset{\boldsymbol{\pi}\in \{0,1\}}{\text{minimize}} \quad  \quad
E_{\text{tot}}=\hat{E}+\mathbb{E}_\Phi[E^*],
\end{equation}
where $\mathbb{E}_\Phi[.]$ represents the expectation function with respect to $\Phi$ and $E^*$ can be obtained as follows:
\begin{align}
&E^*=\underset{\boldsymbol{\epsilon} \in \{0,1\},\boldsymbol{\zeta} \in \{0,1\}}{\text{minimize}} \quad  \quad
f(\boldsymbol{\epsilon},\boldsymbol{\zeta})      \label{parL}\\
&\hspace{-0.5cm}\text{subject to:}\nonumber\\
&\hspace{-0.5cm} \sum_{t=1}^{b} E_{d_l}^t-\sum_{t=1}^{b-1} H_{d_l}^t(\varphi_{d_l}^b) \leq  S_{d_l}^{0}, \quad \forall l, \forall b,\label{pconsumingL}\\
&\hspace{-0.5cm}  S_{d_l}^{0}+\sum_{t=1}^{b} H_{d_l}^t(\varphi_{d_l}^b)-\sum_{t=1}^{b-1} E_{d_l}^t \leq  \bar{S},  \forall l, \forall b, \label{pstoringL}\\
&\hspace{-0.5cm}\sum \limits_{i=0}^Z \epsilon^{b}_{d_l}(i)=1, \quad \forall l, \forall b,\label{pepsilonDroneL}\\
&\hspace{-0.5cm}\sum \limits_{l=1}^D \epsilon^{b}_{d_l}(i)\leq 1, \quad \forall i=1,\cdots,Z, \forall b,\label{pepsilonLocationL}\\
&\hspace{-0.5cm}U^b - \sum \limits_{k=1}^M \pi^b_{m_k} U^b_{m_k}- \sum \limits_{l=1}^D \sum \limits_{i=1}^Z \epsilon^b_{d_l}(i) U^b_{d_l,i} \leq \bar{U}_0, \quad \forall  b,      \label{pcapacity0L}\\
&\hspace{-0.5cm}\zeta^{b}_{d_l}(j,i) \leq \epsilon^b_{d_l}(i),\quad \forall l, \forall i, \forall j, \forall b,\label{pzeta1}\\
&\hspace{-0.5cm}\zeta^{b}_{d_l}(j,i) \leq \epsilon^{b-1}_{d_l}(j),\quad \forall l, \forall i, \forall j, \forall b,\label{pzeta2}\\
&\hspace{-0.5cm}\zeta^{b}_{d_l}(j,i) \geq \epsilon^{b-1}_{d_l}(j)+\epsilon^b_{d_l}(i)-1,\quad \forall l,\forall i, \forall j,\forall b,\label{pzeta3}
\end{align}
where $f(\boldsymbol{\epsilon},\boldsymbol{\zeta}) $ is a function of $\boldsymbol{\epsilon}$ and $\boldsymbol{\zeta}$. In this case, $\boldsymbol{\epsilon}$ and $\boldsymbol{\zeta}$ are considered the second stage decision variables.
The solution of the first stage problem can be solved efficiently by evaluating the expectation over $\boldsymbol{\Phi}$, in case the solution of the second stage problem can be obtained in its closed-form expression. However, in most cases, obtaining a closed-form solution may either be impossible or requires the computation of very complicated and intractable expressions. In order to simplify the problem, we propose to discretize the random variables to solve the two stage problem recourse efficiently~\cite{twostage}. This allow the achievement of near optimal solutions for continuous random variables with an accuracy level dependent on the discretization scale.

We assume that the random variables $\varphi_{d_l}^b, \forall l=1,\cdots,D, \forall b=1,\cdots,B$ is discretized to take a set of $W$ possible values. We denote by $\mathcal{W}$ the set that includes all the possible combinations of the RE generation over the drones. Its size is given as $|\mathcal{W}|=W^{DB}$, where $|.|$ denotes the cardinality of a set, and depends on the number of drones $D$ and the number of time slots $B$. We consider that each possibility of $W$ is realized with a probability $\mathcal{P}_w, w=1,\cdots,|\mathcal{W}|$ where $\mathcal{P}_w$ indicates the probability mass function of $\varphi_{d_l}^b$ which can be determined from the discretization process. Therefore, the two stage recourse optimization problem can be formulated as the following large binary linear programming problem:
\begin{align}
&\hspace{-0.5cm}\underset{\underset{\boldsymbol{\zeta}_w \in \{0,1\}}{\boldsymbol{\epsilon}_w \in \{0,1\}, \boldsymbol{\pi} \in \{0,1\},}}{\text{minimize}} \quad  \quad
\hat{E}+\mathbb{E}_\Phi[f(\boldsymbol{\epsilon}_w,\boldsymbol{\zeta}_w)]      \label{sparL}\\
&\hspace{-0.5cm}\text{subject to:}\nonumber\\
&\hspace{-0.5cm} \sum_{t=1}^{b} E_{d_l,w}^t-\sum_{t=1}^{b-1} H_{d_l,w}^t(\varphi_{d_l}^b) \leq  S_{d_l}^{0}, \quad \forall l, \forall b, \forall w, \label{spconsumingL}\\
&\hspace{-0.5cm}  S_{d_l}^{0}+\sum_{t=1}^{b} H_{d_l,w}^t(\varphi_{d_l}^b) -\sum_{t=1}^{b-1} E_{d_l,w}^t \leq  \bar{S},  \forall l, \forall b, \forall w, \label{spstoringL}\\
&\hspace{-0.5cm}\sum \limits_{i=0}^Z \epsilon^{b}_{d_l,w}(i)=1, \quad \forall l, \forall b, \forall w,\label{spepsilonDroneL}\\
&\hspace{-0.5cm}\sum \limits_{l=1}^D \epsilon^{b}_{d_l,w}(i)\leq 1, \quad \forall i=1,\cdots,Z, \forall b, \forall w,\label{spepsilonLocationL}\\
&\hspace{-0.5cm}U^b - \sum \limits_{k=1}^M \pi^b_{m_k} U^b_{m_k}- \sum \limits_{l=1}^D \sum \limits_{i=1}^Z \epsilon^b_{d_l,w}(i) U^b_{d_l,i} \leq \bar{U}_0, \forall  b, \forall w,     \label{spcapacity0L}\\
&\hspace{-0.5cm}\zeta^{b}_{d_l,w}(j,i) \leq \epsilon^b_{d_l,w}(i),\quad \forall l, \forall i, \forall j, \forall b, \forall w,\label{spzeta1}\\
&\hspace{-0.5cm}\zeta^{b}_{d_l,w}(j,i) \leq \epsilon^{b-1}_{d_l,w}(j),\quad \forall l, \forall i, \forall j, \forall b, \forall w,\label{spzeta2}\\
&\hspace{-0.5cm}\zeta^{b}_{d_l,w}(j,i) \geq \epsilon^{b-1}_{d_l,w}(j)+\epsilon^b_{d_l,w}(i)-1,\quad \forall l,\forall i, \forall j,\forall b, \forall w,\label{spzeta3}
\end{align}
where $\mathbb{E}_\Phi[f(\boldsymbol{\epsilon}_w,\boldsymbol{\zeta}_w)]= \sum\limits^{|\mathcal{W}|}_{w=1} \mathcal{P}_w f(\boldsymbol{\epsilon}_w,\boldsymbol{\zeta}_w)$. The optimal solution for the binary linear optimization problem given in \eqref{sparL}-\eqref{spzeta3} can be determined using Gurobi/CVX interface~\cite{Gurobi}.
Notice that this problem becomes very complex compared to the other scenarios as the number of its constraints exponentially scales with the number of possibilities $W^{DB}$.

\subsection{Complexity and Infesability Discussion}
In the previous sections, we have developed generic optimization problems for three cases depending on the level of knowledge about the renewable energy management. The problems are modeled in the form of non-linear integer programming that we linearize to solve them optimally. Linear integer programming is solved optimally using algorithms such as branch and bound and these are integrated in off-the shelf software such as Gurobi/CVX or Cplex.
In Table~\ref{tabComlexity}, we have provided the complexity for all knowledge cases. As discussed before, the zero knowledge case is the least complex of the cases, but it has to be executed $B$ times, i.e., at the beginning of each time slot. On the other hand, the perfect and partial knowledge cases which are more computationally complex need to be executed once. It is known that these binary programming problems are NP-hard and their complexities grow as the size of the problem grows. In practice, for ultra dense networks, low complexity approaches, as the one presented in Section~\ref{Algo}, can be employed to achieve sub-optimal solutions and the present results can be used to evaluate the efficiency of the developed sub-optimal solutions. Otherwise, the macrocell area can be divided into multiple regions, for example, according to the number of its sectors. Hence, the area covered by each sector antenna is managed separately and a number of drones is assigned to each subarea. Another potential solution is to divide the area into subareas with different priority levels that represent, for instance, the QoS per subarea or the users' density and hence, different numbers of drones are assigned to each subarea. In all cases, the huge complexity does not pose a serious concern for the operator since the algorithms are proactive and do not have to be repeatedly executed.
In Section~\ref{Algo}, we propose a heuristic relaxed solution based on subgradient algorithm that reaches near optimal solution with low complexity.

The proposed solutions are centralized and the decision is made by the macrocell BS that is managing the rest of the BSs. The macrocell BS is responsible in gathering the information about the network traffic within its cell and makes the right decision using the proposed solutions to determine the active BSs and the drones to be used in the next time slots according to the energy availability at the drones' batteries. The decision is made at the beginning of each time slot for the zero knowledge case while it is made once at the beginning of the time horizon for the perfect and partial knowledge cases. The duration of the overhead exchange is very small compared to the length of the time slots and the time horizon, and it can be therefore neglected.
{\small
\begin{table}[h!]
\centering
\caption{Problem Complexity}
\label{tabComlexity}
\addtolength{\tabcolsep}{-4pt}\begin{tabular}{|c||c|c|c|}
\hline
\textbf{Variable/Constraint} & \textbf{Zero knowl.} & \textbf{Perfect knowl.} & \textbf{Partial knowl.}\\ \hline \hline
$\boldsymbol{\pi}$ & $M$ & $MB$ & $MB| \mathcal W| $    \\ \hline
$\boldsymbol{\epsilon}$ & $D$  & $DB$ & $DB| \mathcal W| $\\ \hline
$\boldsymbol{\zeta}$ & $-$ & $BD(Z+1)^2$ & $BD(Z+1)^2| \mathcal W|$  \\ \hline\hline
\eqref{consuming}, \eqref{consumingL}, or \eqref{spconsumingL} & $D$& $DB$ & $DB|\mathcal W|$ \\ \hline
\eqref{storing}, \eqref{storingL}, or\eqref{spstoringL} & $D$& $DB$ & $DB|\mathcal W|$ \\ \hline
\eqref{epsilonDrone}, \eqref{spepsilonDroneL}, or \eqref{spepsilonDroneL} & $D$& $DB$ & $DB|\mathcal W|$ \\ \hline
\eqref{epsilonLocation}, \eqref{epsilonLocationL}, or \eqref{spepsilonLocationL}& $Z$ & $ZB$ & $DB|\mathcal W|$ \\ \hline
 \eqref{capacity0}, \eqref{capacity0L}, or \eqref{spcapacity0L} & $1$ & $B$ & $B|\mathcal W|$ \\ \hline
Linearization constraints & & & \\
\eqref{zeta1}, \eqref{zeta2}, and \eqref{zeta3} & $-$ & $3DB(Z+1)^2$ & $3DB(Z+1)^2|\mathcal W|$ \\
or \eqref{spzeta1}, \eqref{spzeta2}, and \eqref{spzeta3} & & & \\\hline
\end{tabular} \vspace{-.2cm}
\end{table}
}

In this framework, the overloading risk occurs when the optimization problem is infeasible. In other words, the solver is not able to satisfy the constraints~\eqref{capacity0}, \eqref{capacity0L}, and \eqref{pcapacity0L} are not satisfied for the zero, perfect, and partial knowledge cases, respectively at least for one of the time slots. Hence,
\begin{equation}
\exists	 \,b \mbox{ such that } U^b \geq   \bar{U}_0+\sum\limits_{k=1}^M \pi^b_{m_k} U^b_{m_k}+\sum\limits_{l=1}^D \sum \limits_{i=1}^Z \epsilon^b_{d_l}(i) U^b_{d_l,i}.
\end{equation}
The infeasibility condition occurs when the cell capacity constraint of the macrocell BS is violated since the system gives priority to small cells (i.e., DBS and MBS) to serve users. The cell capacity constraint is not satisfied only if the radiated power of the macrocell BS expressed in~\eqref{radiated_power} exceeds its power budget. To cope with overloading risk, the operator is either required to increase the number of MBSs or DBS. In our results, we have also shown that, for the same number of DBSs, powering them with solar panels can help in reducing the overloading risk by reducing the need of extra to and fro trips to the charging station.

\subsection{Relaxed solution}\label{Algo}
In this section, we propose a heuristic solution based on the relaxation of the binary variables to solve our optimization problems with lower complexity. In the first step, we convert the optimization problem into linear programming problem by relaxing the optimization variables $0 \leq \boldsymbol{\zeta},\boldsymbol{\epsilon}, \boldsymbol{\pi} \leq 1$. Now, the linear programming can be formulated as follows:
\begin{align}
&\underset{\boldsymbol{x}}{\text{minimize}} \quad c^T \boldsymbol{x}    \label{Li0}\\
&\text{subject to:}\nonumber\\
& A\boldsymbol{x} \leq A_0,\label{Li1}
\end{align}
where $\boldsymbol{x}$ represents the vector containing the relaxed decision variables while the elements of the vectors $\boldsymbol{c}$ and $A_0$ are deduced from the objective function and constraints, respectively.

In the second step, we employ the subgradient method to find the optimal Lagrangian multipliers ($\boldsymbol{\lambda}$) of this linear programming problem (see \cite{alsharoaTCCN} for more details). Hence, to obtain the solution, we can start with any initial values for the different Lagrangian multipliers and evaluate the optimal solution. We then update the Lagrangian multipliers at the next iteration $(r + 1)$ as follows:
\begin{equation}
\boldsymbol{\lambda}^{(r+1)}=\boldsymbol{\lambda}^{(r)}- \boldsymbol{\varpi}^{(r)} \left( A_0- A\boldsymbol{x}\right),
\end{equation}
where $\boldsymbol{\varpi}^{(r)}$ is the updated step size vector in iteration $r+1$ according to the nonsummable diminishing step length policy (see \cite{subgradient} for more details). The updated values of the
optimal solution and the Lagrangian multipliers are repeated until convergence.
In the third step, we reconstruct the binary variables by rounding them, however, if the rounding violates the battery constraint then, we send the corresponding drone(s) to the charging station. For example, if $\epsilon^b_{d_2}(i)=0.7$, then we round it to 1 and verify the solution feasibility, (i.e.,  we send drone $2$ to the charging station if the rounding violates his battery constraints, or we keep it for the whole $T_b$ at position $i$ otherwise.

\section{Numerical Results}\label{Simulations}
In this section, selected numerical results are provided to investigate the benefits of utilizing dynamic DBSs in HetNets. Firstly, the results are investigating the zero knowledge and perfect knowledge cases to show the performance and advantages of our proposed drone-assisted HetNet model. Then, a comparison with the partial knowledge case is performed to evaluate the impact of uncertainty on the system performance. Finally, we compare the performance of the relaxed method with the one of the optimal solution for the different cases.

In this simulation, we consider two user distributions types, fixed and uniform user distributions. For the fixed type, each drone serves the same number of users while in the uniform case the user distributions follow a uniform distribution with fixed sum.

\subsection{Simulation Parameters}

{\small
\begin{table}[t!]
\centering
\caption{\label{tab2} System parameters}
\addtolength{\tabcolsep}{-2pt}\begin{tabular}{|l|c||l|c||l|c|}
\hline
\textbf{Parameter} & \textbf{Value} & \textbf{Parameter} & \textbf{Value}& \textbf{Parameter} & \textbf{Value}\\ \hline \hline
$\lambda$ (m) & 0.125 & $P_{\text{min}} (dBm)$ & -70  &$T_b$ (minute) & $10$   \\ \hline
$\nu_1$ & $9.6$ & $\nu_2$ & $0.29$ & $\xi_{\text{LoS}}$ (dB) & 1 \\ \hline
$\xi_{\text{NLoS}}$ (dB) & 12 & $\alpha_0$ & 4.7 & $\beta_0$ (W) & 130 \\ \hline
$\alpha_m$ & 2.6 & $\beta_m$ (W) & 56 & $\gamma_m$ (W) & 39 \\ \hline
$\alpha_d$ & 4 & $\beta_d$ (W) & 6.8 & $\gamma_d$ (W) & 2.9 \\ \hline
$\bar{S}$ (kJ) & $10$ & $v_d=v_{\text{max}}$ (m/s) & 15 & $m_{\text{tot}}$ (g) & 750 \\ \hline
$r_p$ (cm) & $20$ & $n_p$ & 4 & $P_s$ (W) & 0.5\\ \hline
$P_{ch}$ (W) & $10$ & $\eta$ & 0.6 & $\bar{U}_0$ & 130\\ \hline
\end{tabular}\vspace{-0.5cm}
\end{table}
}

We assume a HetNet consisting of one macrocell BS with radius of one km, four MBSs ($M=4$) with a coverage of 250 meters, and six identical drones ($D=6$), unless otherwise stated, that can potentially be placed in sixteen different locations ($Z=16$) in addition to the charging station location. We consider that these $16$ locations have the same altitude $h_i=60$ meters, $\forall i=1,\dots,16$ and that each drone has a coverage of $150$ meters to meet the minimum required receiving power $P_\text{min}=-70$ dBm. The $Z+1$ pre-planned locations are indicated as depicted in Fig~\ref{fig1}. We assume that the drones are initially charged with $S^0_{d_l}=6$ kJ of energy and placed at the charging station. The average received amount of photovoltaic power $\phi^b_{d_l}$ is assumed to be generated following a Gamma distribution with shape and scale parameters equal to 1 and 2, respectively. We assume that $U^b=140, \forall b=1,\cdots,B$ users exist within the macrocell BS unless otherwise stated. In case of cell overlap between the MBS and an active DBS, we assume that the drone has the priority in serving the users in the intersection region once deployed. In Table~\ref{tab2}, we present the values of the remaining parameters used in the simulations~\cite{EARTH,Dpower_model}.

\subsection{System Performance}

\begin{figure}[h!]
  \centerline{\includegraphics[width=3.5in]{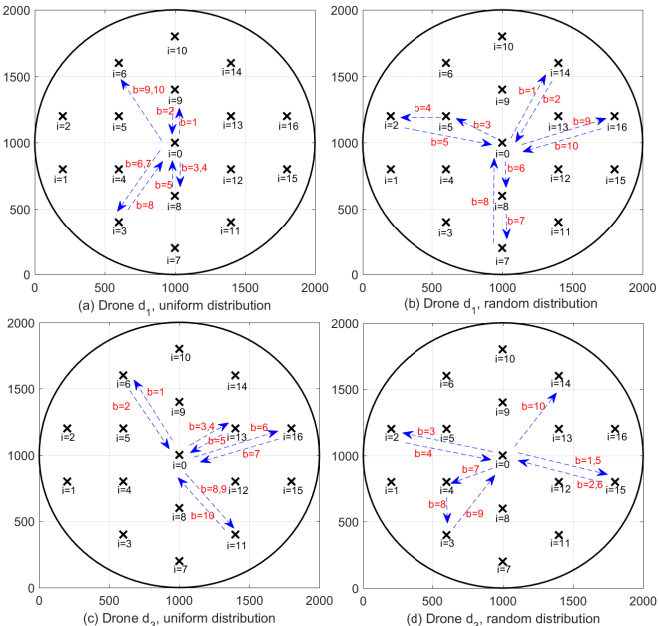}}\vspace{-0.2cm}
   \caption{\, The behavior of two drones, drone $d_1$ (a,b), drone $d_3$ (c,d) for different user distributions with $D=6$.\vspace{-0.4cm}}\label{fig1}
\end{figure}
In Fig.~\ref{fig1}, we start by investigating the behavior of two randomly selected drones, drone $d_1$ and drone $d_3$ respectively, for two different user distributions but for the same number of users and RE generation per drone and time slot. In Fig.~\ref{fig1}(a,c), we consider a uniform user distribution and hence, if a drone is placed in a location $i\neq 0$, it will serve, on average, exactly the same number of users as another drone placed in another location $j\neq i$. In Fig.~\ref{fig1}(b,d), another non-uniform distribution is considered and hence, the number of users to be served differs from a location to another. It is shown that with the uniform distribution, once the drone is sent to a location $i$ then, it has two possibilities for the next slot, either to stay at the same location (e.g., $d_1$ during $b=3,4$) if it has enough energy, otherwise, it returns back to the charging station (e.g., $d_1$ during $b=2$). On the other hand, with the random distribution, the drone can go from one location to another to serve the users without passing by the charging station. For instance, $d_1$ goes to $i=5$ in $b=3$, then moves to $i=2$. It is also worth to note that the drones avoid long distance trip when selecting the locations unless they are forced to do it due to high user density in these locations (e.g., $d_3$ with random distribution moves to $i=2, 15$ during $b=3,1,5$).

\begin{figure}[h!]
  \centerline{\includegraphics[width=3.5in]{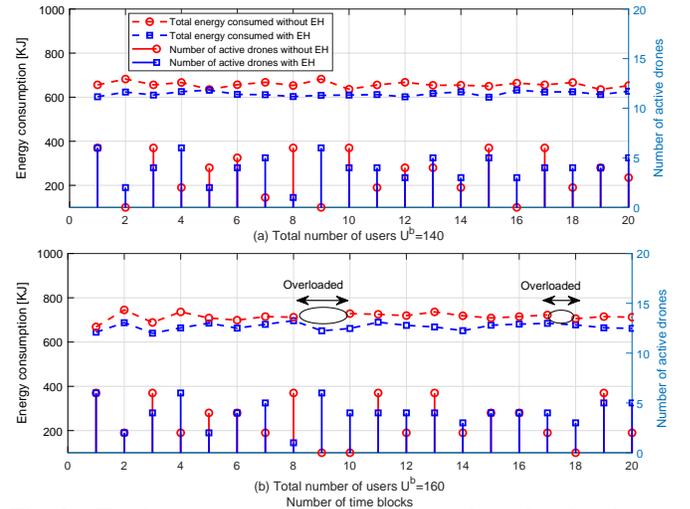}}\vspace{-0.2cm}
   \caption{\, Total energy consumption and number of active drones during the trial period for (a) $U^b=140$ and (b) $U^b=160, \forall b$.\vspace{-0.4cm}}\label{fig2}
\end{figure}

\begin{figure}[h!]
  \centerline{\includegraphics[width=2.8in]{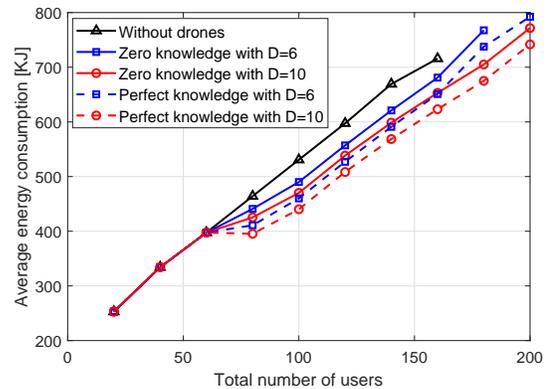}}\vspace{-0.2cm}
   \caption{\, Comparison between the zero knowledge and perfect knowledge cases for different values of $D$.\vspace{-0.2cm}}\label{fig3}
\end{figure}

In Fig.~\ref{fig2}, we plot the energy consumption and number of active drones per time slot for $B=20$ and different number of users uniformly distributed (i.e., $U^b=140$ and $U^b=160, \forall b$). This figure investigates the impact of RE for two cases: 1) when the drones are supported by solar panels and 2) when the drones are charged by the central station only. Please note that, in Fig. 3, the blue lines represent with help of RE case and red lines represent without the help of RE.
It is shown that EH does not only help in reducing the total fossil fuel consumption by the network, but it also it helps in avoiding (or decreasing) the overloading risk. Note that overloading happens when at least one user is not covered by service (i.e., when not all users are simultaneously served). Note that the overloading problem can happen because of at least one of three reasons: 1) not enough number of drones, 2) high demands, and 3) not enough energy of the drones. In the latter reason, the drones need to go back to the charging station more frequently.

Indeed, the ellipses in this figure are correspond to time periods where the system is not able to satisfy the user demand using the current infrastructure. For instance, when the number of users in the network is relatively large (e.g., $U^b$=160), two outage periods are detected $b=\{9,10\}$ and $b= 18$. These outages are due to two reasons: Firstly, the non-EH drones need to go to the central station to charge their batteries more frequently than the solar-powered drones. Secondly, the EH drones can harvest energy when flying and serving users which contributes to the increase of their battery levels and hence, get more flexibility to move to other locations without passing by the charging station. This is deduced from the number of active drones of each case which is higher for the EH case.

for a high number of users, the case without drone is not able to satisfy the user need during the whole period as highlighted by the ellipses that we added in the figures. The ellipses corresponds to time periods where the system is not able to satisfy the user demand using the current infrastructure. However, for the same scenario and by adding drones, we were able to serve all users and maintain network stability.

\begin{figure*}
\vspace{-2mm}
\begin{tabular}{l}
\begin{minipage}[t!]{0.9775\textwidth}
\begin{center}
\includegraphics[width=6in]{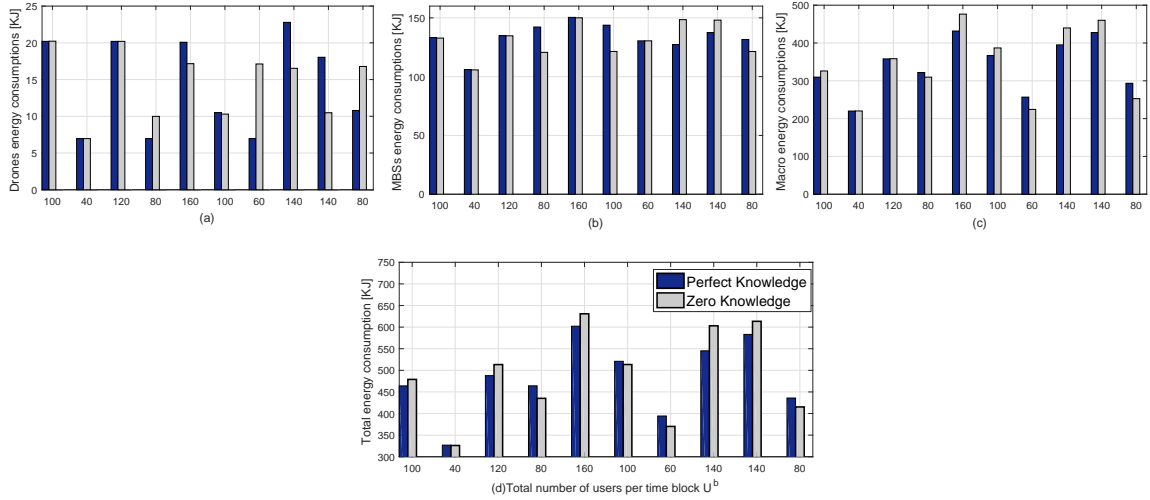}
\caption{\, Comparison between zero and perfect knowledge cases for $D=4$ and $B=10$.}\label{bigfigure}
\end{center}
\vspace{1pt}
\end{minipage}\\
\end{tabular}
\vspace{-15pt}
\end{figure*}

Fig.~\ref{fig3} compares between the zero and perfect knowledge cases presented in Section~\ref{ProblemFormulation} for different number of drones while increasing the total number of users per time slot. It is noticed that increasing the number of drones help in avoiding network outage and reducing the total energy consumption specially when the network becomes more and more congested. For example, in the traditional case without drone, the network becomes overloaded for a number of users higher than $U^b>160$.
Furthermore, Fig. 4 shows the effect of the user densities on the average energy consumptions. We vary the the user density from 25 users (i.e., low density scenario) to 200 users (high density scenario). It can be deduced that for the low density scenarios  both zero and perfect knowledge cases achieve the same performance since the network can be stabalized without needing to the DBS. On the other hand, for high density scenarios EH-DBSs are needed to reduce the overloaded risk
In addition, the perfect knowledge case achieves a more important energy saving due to its efficient management of the harvested energy compared to the zero knowledge case. Nevertheless, the achieved performance of the zero knowledge case follows the same trend of the one of the perfect knowledge case.

In Fig.~\ref{bigfigure} and Table~\ref{Tabbig}, we investigate another scenario for time varied number of users over $B=10$ time slots and with $D=4$ drones and $M=4$ MBSs. The behavior of the drones and the statuses of MBSs are illustrated for each time slot using the zero and perfect knowledge cases.
Fig.~\ref{bigfigure}(a)-(c) plot the total energy consumption of the drones, MBSs, and macrocell BS, respectively. Also, the total energy consumption per time slot is presented in Fig.~\ref{bigfigure}(d).
It can be noticed that, in general, activating the MBSs and/or DBSs essentially depends on the traffic and drones' battery level. Fig.~\ref{bigfigure} also, shows the advantages of using the MBSs along with the drones in order to reduce the macrocell BS energy, thus, reduce the total consumed energy. For example, although the macrocell BS can handle all the users during $b=3$ (i.e., $U^3=120$), the optimization suggests to activate 3 MBSs and 4 drones for both zero and perfect knowledge cases in order to reduce the total transmit power of the macrocell BS and hence, the total energy consumption.
Another important notice can be deduced from Table~\ref{Tabbig}, although the number of users during $b=4$ (i.e., $U^b=80$) is greater than the number of users during $b=7$ (i.e., $U^b=60$), we activate only one drone during $b=4$ instead of 3 drones during $b=7$. This is can be justified by the fact that since the network was more congested during $b=3$ compared to $b=6$ where more drones were sent then, due to the drones' battery limitation, one drone is activated during $b=4$.

It is also worth to note that there is a kind of alternation between the activation of MBSs and the drones' deployment. If the network is partially congested, we notice that the system decides either to deploy drones or activate MBSs depending on the battery levels. For example, for the perfect knowledge, during $b=7$, 3 MBSs are activated while no drone is used. However, during $b=8$, 2 MBSs are turned off and all drones are employed.

\begin{table}[t!]
\begin{center}
\caption{\, Drones and MBSs status during multiple time slots}
\label{Tabbig}
\begin{tabular}{|c|c|c|c|c|c||c|c|c|c|}%
\hline
   \textbf{} & Number of  &\multicolumn{4}{|c||}{Active MBSs}  & \multicolumn{4}{|c|}{Active drones}\\
  \cline{3-10}
  \textbf{}  & users per $b$   & $m_1$ & $m_2$ &$m_3$ &$m_4$ &$d_1$ &$d_2$ &$d_3$ &$d_4$ \\
  \cline{1-10}
  \multirow{10}{*}{\begin{turn}{90}\textbf{Perfect knowledge case}\end{turn}}  & $U^1=100$   & $\times$ & - &$\times$ &$\times$ &$\times$ &$\times$ &$\times$ &$\times$ \\
  \cline{2-10}
  \textbf{}  & $U^2=40$   & $\times$ & $\times$ &- &- & -&- &- & -\\
  \cline{2-10}
  \textbf{}  & $U^3=120$   &  -& $\times$ &$\times$ &$\times$ &$\times$ &$\times$ &$\times$ &$\times$ \\
  \cline{2-10}
  \textbf{}  & $U^4=80$   & $\times$ & $\times$ &$\times$ &$\times$ &- & -&- &- \\
  \cline{2-10}
  {}  & $U^5=160$   & $\times$ & $\times$ &$\times$ &$\times$ &$\times$ &$\times$ &$\times$ &$\times$   \\
  \cline{2-10}
  \textbf{}  & $U^6=100$   & $\times$ & $\times$ &$\times$ &$\times$ & -&- & -& $\times$ \\
  \cline{2-10}
  \textbf{}  & $U^7=60$   & - & $\times$ &$\times$ &$\times$ &- & -& -& -\\
  \cline{2-10}
  \textbf{}  & $U^8=140$   & $\times$ & - &$\times$ & -&$\times$ &$\times$ &$\times$ &$\times$ \\
  \cline{2-10}
  \textbf{}  & $U^9=140$   & $\times$ & - &$\times$ &$\times$ &$\times$ &$\times$ & -&$\times$ \\
  \cline{2-10}
  \textbf{}  & $U^{10}=80$   & - & $\times$ &$\times$ &$\times$ & -&- &$\times$ &- \\
\hline\hline
  \multirow{10}{*}{\begin{turn}{90}\textbf{Zero knowledge case}\end{turn}}  & $U^1=100$   & $\times$ & - &$\times$ & $\times$ &$\times$ &$\times$ &$\times$ &$\times$ \\
  \cline{2-10}
  \textbf{}  & $U^2=40$   & -& - & -&$\times$ &- &- &- &- \\
  \cline{2-10}
  \textbf{}  & $U^3=120$   & $\times$ & $\times$ &$\times$ &- &$\times$ &$\times$ &$\times$ &$\times$ \\
  \cline{2-10}
  \textbf{}  & $U^4=80$   & $\times$ & $\times$ &- &- & $\times$& -& -&- \\
  \cline{2-10}
  {}  & $U^5=160$   & $\times$ & $\times$ & $\times$& $\times$& -& $\times$ & $\times$ & $\times$ \\
  \cline{2-10}
  \textbf{}  & $U^6=100$   & $\times$ & $\times$ &- &- & $\times$ &- & -& - \\
  \cline{2-10}
  \textbf{}  & $U^7=60$   & $\times$ & - &$\times$ &$\times$ &- & $\times$& $\times$& $\times$\\
  \cline{2-10}
  \textbf{}  & $U^8=140$   & $\times$ & $\times$ &$\times$ & $\times$ &$\times$ &$\times$ &$\times$& -\\
  \cline{2-10}
  \textbf{}  & $U^9=140$   & $\times$ & $\times$ &$\times$ &$\times$ &- &- & -&$\times$ \\
  \cline{2-10}
  \textbf{}  & $U^{10}=80$   & $\times$ & $\times$ &- &- & $\times$& $\times$&$\times$ &-\\
  \hline
  \end{tabular}
\end{center} \vspace{-0.5cm}
\end{table}

\begin{figure}[h!]
  \centerline{\includegraphics[width=3.5in]{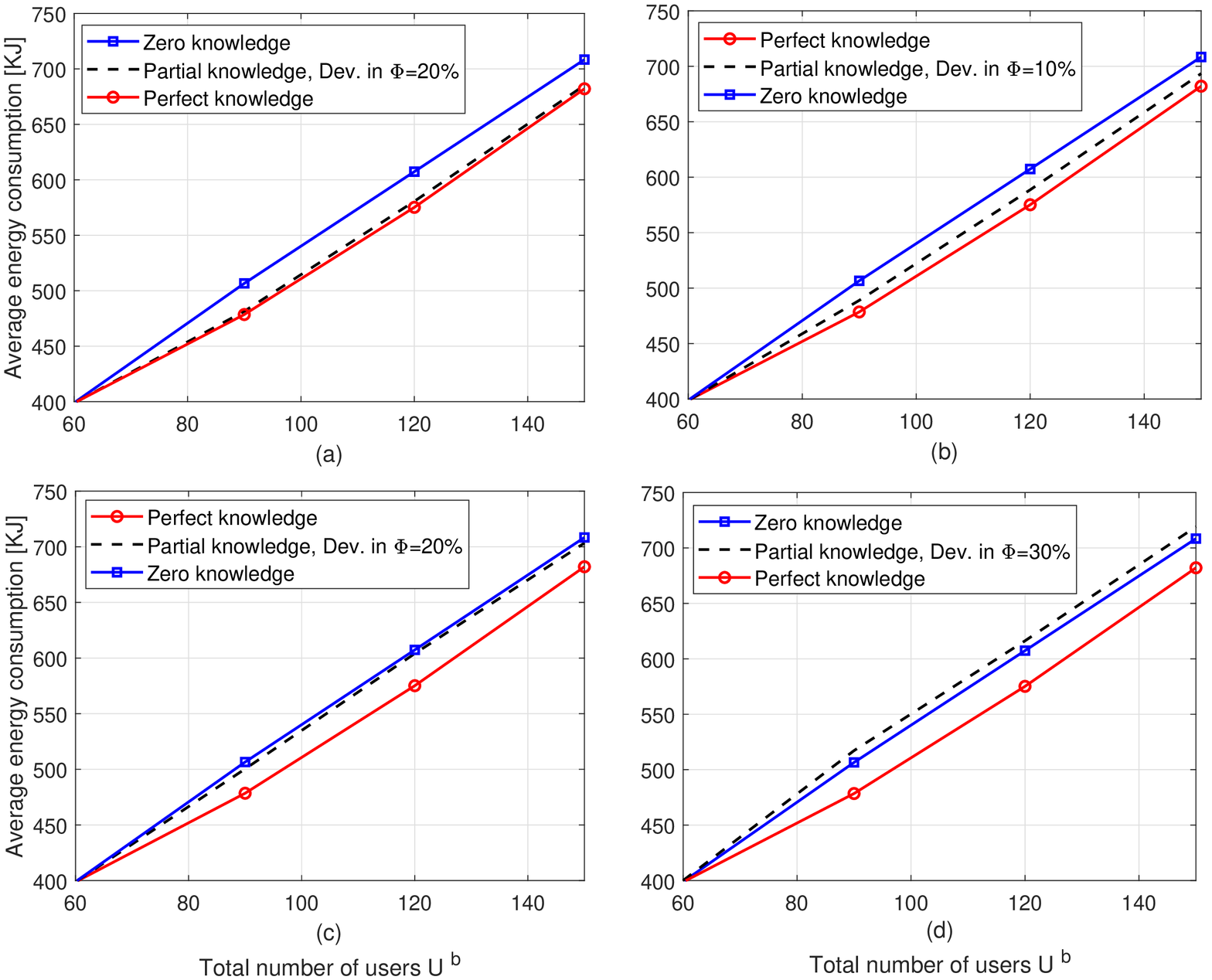}}\vspace{-0.2cm}
   \caption{\, Comparison of the average energy consumption per time slot versus the total number of users for $D=3$ with different deviation values in $\Phi$.\vspace{-0.4cm}
   }\label{fig4}
\end{figure}

\begin{figure}[h!]
  \centerline{\includegraphics[width=2.7in]{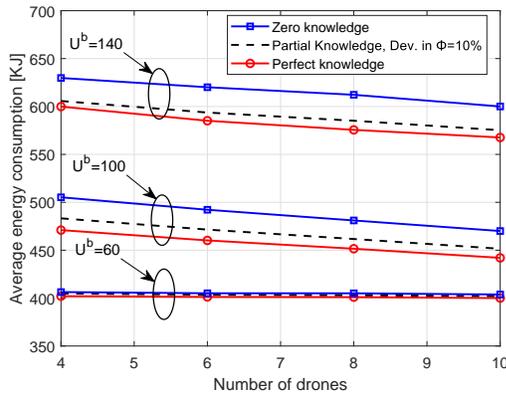}}\vspace{-0.2cm}
   \caption{Average energy consumption versus number of drones for different user traffic scenarios.\vspace{-0.4cm}}\label{energy_vs_drones}
\end{figure}

\begin{figure*}
\vspace{-2mm}
\begin{tabular}{l}
\begin{minipage}[t!]{0.9775\textwidth}
\begin{center}
\includegraphics[width=6in]{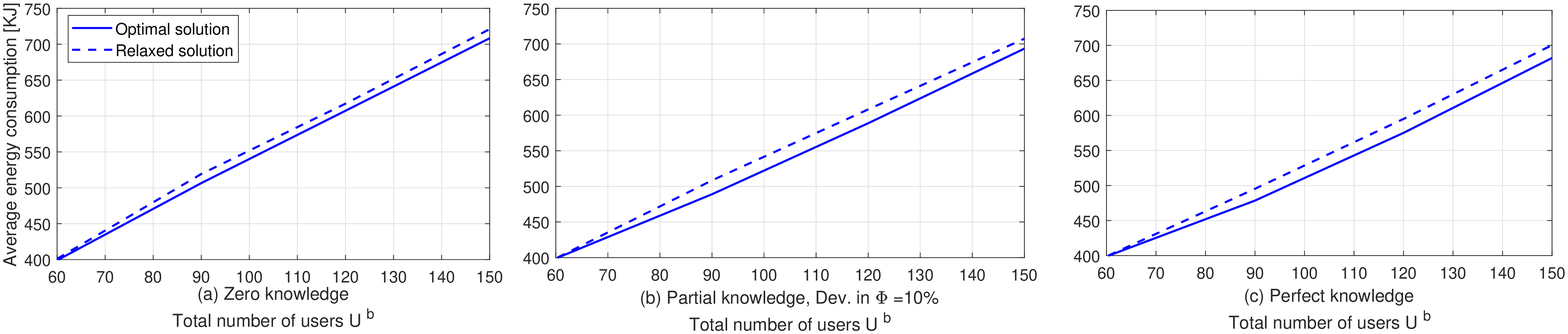}
\caption{\, Comparison between the optimal and relaxed solutions. \vspace{-0.4cm}}\label{fig55}
\end{center}
\vspace{1pt}
\end{minipage}\\
\end{tabular}
\vspace{-15pt}
\end{figure*}

On the other hand, it can be shown that the prefect knowledge case achieves better performance than the one of the zero knowledge case by managing the available resource more efficiently such as drones' available batteries. For example, as shown in Table~\ref{Tabbig}, during $b=7$ (i.e., $U^7=160$), the perfect knowledge case keeps all the drones in the charging station in order to charge the batteries and hence, it becomes possible to send most of them during the next two time slots (i.e., $d_1,d_2,d_3$, and $d_4$ are active during $b=8$ and $d_1,d_2$, and $d_4$ are active during $b=9$) where the network is more congested, i.e., $U^8=U^9=140$ as shown in Table~\ref{Tabbig}.
As shown in Fig.~\ref{bigfigure}, although it consumes more energy than the zero knowledge case, which is around $20$ kJ, when $b=7$, the perfect knowledge case saves more energy, which is around $75$ kJ, during the next two time slots $b=8$ and $b=9$.

Fig.~\ref{fig4} plots the average energy consumption per time slot $b$ (i.e., $E_{\text{tot}}/B$) versus the total number of users for $D=3$. The results compare between the different RE knowledge cases: zero, perfect, and partial knowledge cases, where different deviation values from the mean of RE generations $\Phi$ are considered in the partial knowledge case. In other words, we consider a discrete decision variable with 2 possibilities where $\tilde{\Phi} \in\{-x\% \bar{\Phi}, x\%\bar{\Phi}\}$. In the figure, we set $x=\{5,10,20,30\}$. The obtained results confirm that the perfect knowledge case always achieves the lowest energy consumption compared to the other cases (zero and partial knowledge cases) as it represents the benchmark solution. On the other hand, the partial knowledge case achieves better performance compared to the zero knowledge case and the obtained energy consumption remains close to the perfect case when the uncertainty is relatively small (e.g., $5\%,10\%,20\%$). However, when the uncertainty is relatively large, (e.g., $30\%$), the zero knowledge case outperforms the partial knowledge case since the drones in the latter case can not consume the available power in their batteries efficiently. Indeed, when the uncertainty level becomes high, the stochastic programming solution reduces the risk of failing in an outage scenario (either in terms of battery depletion or in terms of network outage). Therefore, it forces the drones to return to the charging station more frequently than the zero knowledge case. Hence, more MBSs are activated and a major part of the users are served by the macrocell BS. This happens clearly, when the number of users is relatively large.

Fig.~\ref{energy_vs_drones} plots the average energy consumption versus the number of drones for the three drone management knowledge cases (i.e., zero, partial, perfect). It also considers three user density scenarios (low traffic, moderate traffic, and high traffic). It can be noticed that for the low traffic scenario (i.e., $U^b=60$ users), almost all the management knowledge cases perform the same. This can be justified by the fact that the network can handle the low number of users without need of the drones on average. While for moderate or high traffic scenarios (i.e., $U^b=\{100,140\}$ users), the drones are needed in order to enhance the performance. Also, from this figure, we can conclude that using more drones can reduce the average energy consumption significantly. For instance, using 10 drones instead of 4 drones can reduce the average energy consumption by around 8\% by consuming around 450 KJ instead of  480 KJ using partial knowledge case.

Fig.~\ref{fig55} compares the performances of the relaxed methods to those of the optimal solutions  solution. It is shown that the relaxed solution achieves a close average energy consumption compared to the one of optimal solution with a lower complexity. Also, it can be noticed that for the zero knowledge case, the gap between optimal and relaxed solutions is smaller that the other knowledge cases and this can justified by the fact that the zero knowledge case requires the optimization of a smaller number of variables in comparison with the other cases. It is also worth to note that the gap grows with the increase of the congestion level of the network.

\section{Conclusions}\label{Conclusions}

In this paper, we proposed an energy management framework for cellular heterogeneous networks assisted by solar-powered drone small cells. An integer linear programming problem is formulated in order to minimize the total energy consumption of the networks over a time-slotted period while maintaining the network coverage and connectivity. Multiple drone base stations are optimally placed in order to support overloaded cells while taking into account their photovoltaic energy generation and battery capacity. In order to deal with the uncertainty in the renewable energy generation, two cases are investigated in our analysis. The first case, identified as the zero knowledge case, manages the system time slot per time slot without considering future renewable energy generation. The second case exploits the partial knowledge about future renewable energy generation and devises a pre-planned network management while considering the level of uncertainty in its optimization. These two cases are compared to a benchmark case assuming perfect knowledge of future renewable energy generation, i.e., zero uncertainty.

Through several numerical results, we investigated the behavior of the dynamic drones as well as the ON/OFF switching operation applied to the micro cell BSs for different scenarios. Our results show the notable impacts of employing dynamic drones mainly during peak-hour periods in ensuring connectivity and supporting overloaded cells while minimizing the energy consumption.
The use of the drones is not only for saving energy but also for supporting ground communications and help in reducing the overloading risk.
As expected the perfect knowledge case outperforms the other cases which provides close solutions for low levels of uncertainty. However, for high uncertainty level, the partial knowledge case will be more risk-aware and generates safer solutions to avoid battery depletion and network outage. The results obtained via the relaxation technique allow the achievement of slightly suboptimal solutions but with significantly lower complexity. The gap varies according to the used method and the congestion level of the network.

This paper reinforces the trend of employing aerial base stations to support next-generation cellular networks for different use-cases. This includes, not only the maintenance of connectivity during peak-hours and the replacement of damaged ground infrastructure under the context of public safety communications, but also the continuous support of the ground infrastructure while also supporting greener communications. In the latter use-case, we have shown that an optimized management of all the components of HetNets supported by DBSs allows more power-hungry devices to be turned off, temporarily replace them by less power-hungry devices that are able to offer better channel quality, which leads to additional gain in terms of energy consumption. The operation time of the aerial BSs can be further enhanced by employing renewable energy sources. This reduces the trips of the drones to the charging station for environment-friendly operators.

\bibliographystyle{IEEEtran}
\bibliography{2019J_TMC}

\begin{thebibliography}{10}
\providecommand{\url}[1]{#1}
\csname url@samestyle\endcsname
\providecommand{\newblock}{\relax}
\providecommand{\bibinfo}[2]{#2}
\providecommand{\BIBentrySTDinterwordspacing}{\spaceskip=0pt\relax}
\providecommand{\BIBentryALTinterwordstretchfactor}{4}
\providecommand{\BIBentryALTinterwordspacing}{\spaceskip=\fontdimen2\font plus
\BIBentryALTinterwordstretchfactor\fontdimen3\font minus
  \fontdimen4\font\relax}
\providecommand{\BIBforeignlanguage}[2]{{%
\expandafter\ifx\csname l@#1\endcsname\relax
\typeout{** WARNING: IEEEtran.bst: No hyphenation pattern has been}%
\typeout{** loaded for the language `#1'. Using the pattern for}%
\typeout{** the default language instead.}%
\else
\language=\csname l@#1\endcsname
\fi
#2}}
\providecommand{\BIBdecl}{\relax}
\BIBdecl

\bibitem{alsharoaWCNC}
A.~Alsharoa, H.~Ghazzai, A.~Kadri, and A.~E. Kamal, ``Energy management in
  cellular hetnets assisted by solar powered drone small cells,'' in \emph{2017
  IEEE Wireless Communications and Networking Conference (WCNC), San Francisco,
  CA, USA}, Mar. 2017, pp. 1--6.

\bibitem{surv1}
S.~Hayat, E.~Yanmaz, and R.~Muzaffar, ``Survey on unmanned aerial vehicle
  networks for civil applications: {A} communications viewpoint,'' \emph{IEEE
  Communications Surveys Tutorials}, vol.~18, no.~4, pp. 2624--2661,
  Fourthquarter 2016.

\bibitem{surv2}
L.~Gupta, R.~Jain, and G.~Vaszkun, ``Survey of important issues in uav
  communication networks,'' \emph{IEEE Communications Surveys Tutorials},
  vol.~18, no.~2, Secondquarter 2016.

\bibitem{DBSint1}
{I. Bucaille et al.}, ``Rapidly deployable network for tactical applications:
  {A}erial base station with opportunistic links for unattended and temporary
  events absolute example,'' in \emph{IEEE Military Communications Conference
  (MILCOM)}, Nov. 2013, pp. 1116--1120.

\bibitem{Att1}
{AT\&T/Qualcomm}, ``Qualcomm and {AT\&T} to trial drones on cellular network to
  accelerate wide-scale deployment,''
  \url{http://about.att.com/story/qualcomm_and_att_to_trial_drones_on_cellular_network.html}.

\bibitem{Att2}
{AT\&T/Intel}, ``{AT\&T} and {I}ntel to test drones on {LTE} network,''
  \url{http://about.att.com/story/att_and_intel_to_test_drones_on_lte_network.html}.

\bibitem{Nokia_ref}
{Nokia}, ``Nokia showcases power of drones and {LTE} connectivity for public
  safety at {D4G} award event in {D}ubai,''
  \url{https://www.nokia.com/en_int/news/releases/2017/02/17/nokia-showcases-power-of-drones-and-lte-connectivity-for-public-safe-}{\\\url{ty-at-d4g-award-event-in-dubai}},.

\bibitem{pathdeployment1}
Y.~Zeng, R.~Zhang, and T.~J. Lim, ``Wireless communications with unmanned
  aerial vehicles: {O}pportunities and challenges,'' \emph{IEEE Communications
  Magazine}, vol.~54, no.~5, pp. 36--42, May 2016.

\bibitem{pathdeployment3}
T.~Schouwenaars, B.~DeMoor, E.~Feron, and J.~How, ``Mixed integer programming
  for multi-vehicle path planning,'' in \emph{Proc. of the European Controls
  Conference, Orlando, Florida, USA}, Dec. 2011, p. 2603–2608.

\bibitem{int1}
J.~Xu and R.~Zhang, ``Throughput optimal policies for energy harvesting
  wireless transmitters with non-ideal circuit power,'' \emph{IEEE Journal on
  Slected Areas in Communications}, vol. 32, no. 2, pp. 322--332, Feb. 2014.

\bibitem{RFA2}
H.~Tabassum, E.~Hossain, A.~Ogundipe, and D.~I. Kim, ``Wireless-powered
  cellular networks: {K}ey challenges and solution techniques,'' \emph{IEEE
  Communications Magazine}, vol. 53, no. 6, pp. 63--71, June 2015.

\bibitem{UlukusEH}
S.~Ulukus, A.~Yener, E.~Erkip, O.~Simeone, M.~Zorzi, P.~Grover, and K.~Huang,
  ``Energy harvesting wireless communications: A review of recent advances,''
  \emph{IEEE Journal on Selected Areas in Communications}, vol. 33, no. 3, pp.
  360--381, Mar. 2015.

\bibitem{GCom}
Z.~Hasan, H.~Boostanimehr, and V.~Bhargava, ``Green cellular networks: {A}
  survey, some research issues and challenges,'' \emph{IEEE Communications
  Surveys \& Tutorials}, vol. 13, no. 4, pp. 524--540, Fourth 2011.

\bibitem{RE}
V.~Raghunathan, S.~Ganeriwal, and M.~Srivastava, ``Emerging techniques for long
  lived wireless sensor networks,'' \emph{IEEE Communications Magazine}, vol.
  44, no. 4, pp. 108--114, Apr. 2006.

\bibitem{H1}
I.~Bor-Yaliniz and H.~Yanikomeroglu, ``The new frontier in {RAN} heterogeneity:
  {M}ulti-tier drone-cells,'' \emph{IEEE Communications Magazine}, vol.~54,
  no.~11, pp. 48--55, Nov. 2016.

\bibitem{relaydrone}
S.~Rohde and C.~Wietfeld, ``Interference aware positioning of aerial relays for
  cell overload and outage compensation,'' in \emph{IEEE Vehicular Technology
  Conference (VTC Fall), Quebec, QC, Canada}, Sept. 2012, pp. 1--5.

\bibitem{relaydrone2}
P.~Zhan, K.~Yu, and A.~L. Swindlehurst, ``Wireless relay communications using
  ann unmanned aerial vehicle,'' in \emph{IEEE Workshop on Signal Processing
  Advances in Wireless Communications, Cannes, France}, July 2006, pp. 1--5.

\bibitem{11}
J.~Košmerl and A.~Vilhar, ``Base stations placement optimization in wireless
  networks for emergency communications,'' pp. 200--205, June 2014.

\bibitem{H2}
{E. Kalantari, I. Bor-Yaliniz, A. Yongacoglu, and H. Yanikomeroglu}, ``User
  association and bandwidth allocation for terrestrial and aerial base stations
  with backhaul considerations,'' \emph{https://arxiv.org/pdf/1709.07356.pdf},
  Sept. 2017.

\bibitem{14}
Z.~Han, A.~L. Swindlehurst, and K.~J.~R. Liu, ``Optimization of {MANET}
  connectivity via smart deployment/movement of unmanned air vehicles,''
  \emph{IEEE Transactions on Vehicular Technology}, vol.~58, no.~7, pp.
  3533--3546, Sept. 2009.

\bibitem{connectivity2}
I.~Bekmezci, M.~Ermis, and S.~Kaplan, ``Connected multi {UAV} task planning for
  flying ad hoc networks,'' in \emph{Proc. of the IEEE International Black Sea
  Conference on Communications and Networking {(BlackSeaCom)}}, May 2014, pp.
  28--32.

\bibitem{safepath1}
J.~Scherer and B.~Rinner, ``Persistent multi-{UAV} surveillance with energy and
  communication constraints,'' in \emph{Proc. of the IEEE International
  Conference on Automation Science and Engineering (CASE)}, Aug. 2016, pp.
  1225--1230.

\bibitem{safepath2}
T.~Schouwenaars, J.~How, and E.~Feron, ``Receding horizon path planning with
  implicit safety guarantees,'' in \emph{Proc. of the American Control
  Conference}, June 2004, pp. 5576--5581.

\bibitem{Drone_saad}
M.~Mozaffari, W.~Saad, M.~Bennis, and M.~Debbah, ``Drone small cells in the
  clouds: Design, deployment and performance analysis,'' in \emph{Proc. of the
  IEEE Global Communications Conference (GLOBECOM), San diego, CA, USA}, Dec.
  2015, pp. 1--6.

\bibitem{LOS1}
Q.~Feng, E.~K. Tameh, A.~R. Nix, and J.~McGeehan, ``Modelling the likelihood of
  line-of-sight for air-to-ground radio propagation in urban environments,'' in
  \emph{Proc. of the IEEE Global Communications Conference (GLOBECOM), San
  Francisco, CA, USA}, Nov.- Dec 2006, pp. 1--5.

\bibitem{LOS2}
Q.~Feng, J.~McGeehan, E.~K. Tameh, and A.~R. Nix, ``Path loss models for
  air-to-ground radio channels in urban environments,'' in \emph{Proc. of the
  63rd IEEE Vehicular Technology Conference (VTC Spring), Melbourne, Vic,
  Australia}, May 2006, pp. 2901--2905.

\bibitem{PL1}
A.~Al-Hourani, S.~Kandeepan, and A.~Jamalipour, ``Modeling air-to-ground path
  loss for low altitude platforms in urban environments,'' in \emph{IEEE Global
  Communications Conference (GLOBECOM), Austin, TX, USA}, Dec. 2014, pp.
  2898--2904.

\bibitem{saad2}
M.~Mozaffari, W.~Saad, M.~Bennis, and M.~Debbah, ``Unmanned aerial vehicle with
  underlaid device-to-device communications: Performance and tradeoffs,''
  \emph{IEEE Transactions on Wireless Communications}, vol.~15, no.~6, pp.
  3949--3963, June 2016.

\bibitem{7470932}
S.~Chandrasekharan, K.~Gomez, A.~Al-Hourani, S.~Kandeepan, T.~Rasheed,
  L.~Goratti, L.~Reynaud, D.~Grace, I.~Bucaille, T.~Wirth, and S.~Allsopp,
  ``Designing and implementing future aerial communication networks,''
  \emph{IEEE Communications Magazine}, vol.~54, no.~5, pp. 26--34, May 2016.

\bibitem{7759260}
R.~D'Sa, D.~Jenson, T.~Henderson, J.~Kilian, B.~Schulz, M.~Calvert, T.~Heller,
  and N.~Papanikolopoulos, ``{SUAV:Q - A}n improved design for a transformable
  solar-powered {UAV},'' in \emph{IEEE/RSJ International Conference on
  Intelligent Robots and Systems (IROS 2016)}, Daejeon, South Korea, Oct. 2016,
  pp. 1609--1615.

\bibitem{H3}
P.~Oettershagen, A.~Melzer, T.~Mantel, K.~Rudin, R.~Lotz, D.~Siebenmann,
  S.~Leutenegger, K.~Alexis, and R.~Siegwart, ``A solar-powered hand-launchable
  uav for low-altitude multi-day continuous flight,'' in \emph{Proc. of the
  IEEE International Conference on Robotics and Automation (ICRA)}, May 2015,
  pp. 3986--3993.

\bibitem{H4}
S.~Morton, L.~Scharber, and N.~Papanikolopoulos, ``Solar powered unmanned
  aerial vehicle for continuous flight: Conceptual overview and optimization,''
  in \emph{Proc. of the IEEE International Conference on Robotics and
  Automation}, May 2013, pp. 766--771.

\bibitem{quadcopter}
``{Solar Powered Quadcopter, Xsol-E1.1},'' \url{https://wn.com/solar-drone}.

\bibitem{PowerFilm}
``{PowerFilm® RC Aircraft Series},''
  \url{https://www.solarmade.com/store/product/rc7.2-75}.

\bibitem{RR7}
C.~Liu, B.~Natarajan, and H.~Xia, ``Small cell base station sleep strategies
  for energy efficiency,'' \emph{IEEE Transactions on Vehicular Technology},
  vol. 65, no. 3, pp. 1652--1661, Mar. 2016.

\bibitem{7776901}
H.~Ghazzai, M.~J. Farooq, A.~Alsharoa, E.~Yaacoub, A.~Kadri, and M.~S. Alouini,
  ``Green networking in cellular hetnets: {A} unified radio resource management
  framework with base station {ON/OFF} switching,'' \emph{to appear in IEEE
  Transactions on Vehicular Technology}, 2017.

\bibitem{EARTH}
``Energy efficiency analysis of the reference systems, areas of improvements
  and target breakdown,'' \emph{{E}nergy {A}ware {R}adio and ne{T}work
  tec{H}nologies}, Dec. 2010.

\bibitem{Saaad}
M.~Mozaffari, W.~Saad, M.~Bennis, and M.~Debbah, ``Unmanned aerial vehicle with
  underlaid device-to-device communications: Performance and tradeoffs,''
  \emph{IEEE Transactions on Wireless Communications}, vol.~15, no.~6, pp.
  3949--3963, June 2016.

\bibitem{EH_drone}
H.~Wang, J.~Wang, G.~Ding, L.~Wang, T.~A. Tsiftsis, and P.~K. Sharma,
  ``Resource allocation for energy harvesting-powered d2d communication
  underlaying uav-assisted networks,'' \emph{IEEE Transactions on Green
  Communications and Networking}, vol.~PP, no.~99, pp. 1--1, 2017.

\bibitem{alsharoaICC}
A.~Alsharoa, H.~Ghazzai, M.~Yuksel, A.~Kadri, and A.~Kamal, ``Trajectory
  optimization for multiple {UAVs} acting as wireless relays,'' in \emph{Proc.
  of the IEEE International Conference on Communications (ICC), Kansas city,
  Missori, USA}, May 2018.

\bibitem{PL2}
Y.~W. Y.~Zheng and F.~Meng, ``Modeling and simulation of pathloss and fading
  for air-ground link of haps within a network simulator,'' in \emph{IEEE
  International Conference on Cyber-Enabled Distributed Computing and Knowledge
  Discovery (CyberC), Beijing, China}, Oct. 2013.

\bibitem{PL3}
A.~Al-Hourani, S.~Kandeepan, and S.~Lardner, ``Optimal {LAP} altitude for
  maximum coverage,'' \emph{IEEE Wireless Communications Letters}, vol.~3,
  no.~6, pp. 569--572, Dec. 2014.

\bibitem{Dpower_model}
J.~V. Dries~Hulens and T.~Goedeme, ``How to choose the best embedded processing
  platform for onboard {UAV} image processing,'' in \emph{Proc. of the
  International Joint Conference Computer Vision, Imaging and Computer Graphics
  Theory and Applications (VISIGRAPP), Berlin, Germany}, Mar. 2015.

\bibitem{stochastic_book}
{P. Kall and J. Mayer}, \emph{Stochastic Linear Programming: Models, Theory,
  and Computation. International series in operations research and management
  science}.\hskip 1em plus 0.5em minus 0.4em\relax New York: Springer, 2005.

\bibitem{Gurobi}
``{Gurobi optimizer reference manual},'' 2016. Available [online]:
  http://www.gurobi.com/.

\bibitem{zalloumi}
M.~J. Farooq, H.~Ghazzai, A.~Kadri, H.~ElSawy, and M.~S. Alouini, ``A hybrid
  energy sharing framework for green cellular networks,'' \emph{IEEE
  Transactions on Communications}, vol.~PP, no.~99, pp. 1--1, 2016.

\bibitem{tss}
S.~Kwon, L.~Ntaimo, and N.~Gautam, ``Optimal day-ahead power procurement with
  renewable energy and demand response,'' \emph{IEEE Transactions on Power
  Systems}, vol.~PP, no.~99, pp. 1--1, 2016.

\bibitem{twostage}
C.~I. Fabian and Z.~Szoke, ``Solving two-stage stochastic programming problems
  with level decomposition,'' \emph{Computational Management Science}, vol.~4,
  no.~4, p. 313–353, Oct. 2007.

\bibitem{alsharoaTCCN}
A.~Alsharoa, H.~Ghazzai, E.~Yaacoub, M.~S. Alouini, and A.~E. Kamal, ``Joint
  bandwidth and power allocation for {MIMO} two-way relays-assisted overlay
  cognitive radio systems,'' \emph{IEEE Transactions on Cognitive
  Communications and Networking}, vol.~1, no.~4, pp. 383--393, Dec. 2015.

\bibitem{subgradient}
S.~Boyd and A.~Mutapcic, ``Stochastic {S}ubgradient {M}ethods,'' \emph{{N}otes
  for EE364, Stanford University}, Winter 2006-07.

\end{thebibliography}

\end{document}